\documentclass{aa} 

\usepackage{graphicx}
\usepackage{txfonts}

\usepackage[english]{babel}
\usepackage{amsmath}    
\usepackage{amssymb}    
\usepackage{multicol}        
\usepackage{bm}         
\usepackage{pdflscape}  
\usepackage{flafter}
\usepackage{wrapfig}
\usepackage[export]{adjustbox}
\usepackage{float}
\usepackage{enumitem}
\usepackage{tabularx} 
\usepackage[normalem]{ulem}
\usepackage{subfig}
\usepackage{tablefootnote}

\usepackage[table,xcdraw]{xcolor}
\usepackage{afterpage}

\usepackage[colorlinks=true,linkcolor=blue,citecolor=blue,urlcolor=blue]{hyperref}

\begin{document} 
   
   \title{3D NLTE Sodium abundances in late-type stars} 
   \subtitle{Abundance corrections and synthetic spectra}

   \author{G.~Canocchi\inst{\ref{inst1}} \and
           E.~X.~Wang\inst{\ref{inst1}} \and
           A.~M.~Amarsi\inst{\ref{inst2}} \and
           K.~Lind\inst{\ref{inst1}} \and
           M.~Racca\inst{\ref{inst1}}
          }
    \institute{Department of Astronomy, Stockholm University, AlbaNova University Center, SE-106 91 Stockholm, Sweden\\ \email{gloria.canocchi@gmail.com}\label{inst1} \and 
    Theoretical Astrophysics, Department of Physics and Astronomy, Uppsala University, Box 516, SE-751 20 Uppsala, Sweden\label{inst2} 
    }

 
  \abstract  
   {Neutral sodium is an important tracer of the Galactic chemical evolution, a powerful diagnostic of different stellar populations, and the subject of detailed studies of exoplanet atmospheres via transmission spectroscopy.} 
   {This work aims to study and quantify the errors in stellar analyses of \ion{Na}{I} lines caused by the use of one-dimensional (1D) hydrostatic model atmospheres and the assumption of local thermodynamic equilibrium (LTE).}
   {We studied the line formation of nine \ion{Na}{I} lines in FGK dwarfs and giants via, for the first time, 3D~non-LTE (NLTE) radiative transfer post-processing with the code \texttt{Balder} on 3D radiation hydrodynamic stellar atmospheres from the \texttt{Stagger} grid spanning $T_\mathrm{eff}= 4000$ to $6500$\,K, $\log g = 1.5$ to $5.0$, and [Fe/H]$=-4$ to $+0.5$. 
   }
   {We find that the 3D~NLTE abundance corrections relative to 1D~LTE tend to be negative, and more positive than the corresponding 1D NLTE corrections. This reflects more efficient overionisation in the steeper temperature gradient of the 3D models. The corrections are typically less severe than $-0.1$\,dex for weak lines, but become much larger for saturated lines in low-gravity giants ($\log g \leq 2.0$), even reaching $-0.7$\,dex. 
   However, for the D resonance lines, the 3D~NLTE corrections relative to 1D~LTE become slightly positive at the lowest metallicities in our grid, typically around $+0.05$\,dex at [Fe/H]$=-4$.
   } 
   {We make our 3D~NLTE grid, together with interpolation routines based on radial basis functions and fully connected feedforward neural networks, publicly available.
   This will enable more accurate determination of sodium abundances in present and forthcoming stellar spectroscopic surveys, particularly for metal-poor stars, as well as a better characterisation of the \ion{Na}{I} D lines in exoplanet atmospheres.}

\titlerunning{3D NLTE sodium abundance corrections and synthetic spectra}
\authorrunning{Canocchi G. et al.}
   \maketitle

\section{Introduction}\label{sec: intro}
Neutral sodium (\ion{Na}{I}) is a well-established tracer of the chemical evolution of the Milky Way. Its abundance, derived from spectroscopic analyses of late-type stars, provides key insights into the nucleosynthetic history and chemical enrichment of different Galactic components. Studies of sodium abundances have been conducted across the thin and thick discs (e.g. \citealt{Bensby2014, Owusu2024}), the stellar halo (e.g. \citealt{Nissen2024}), in globular clusters (GCs; e.g. \citealt{Gratton2012, McKenzie2022}), and in open clusters (e.g. \citealt{LoaizaTacuri2023}). The analysis of \ion{Na}{I} abundances has proven particularly valuable for distinguishing between distinct stellar populations, such as accreted and in-situ stars within the Milky Way (e.g. \citealt{Buder2022}), as well as for identifying multiple stellar generations in GCs (e.g. \citealt{Gratton2001}).

Sodium is an odd-$Z$ element, synthesised from various sources, resulting in a complex abundance trend as a function of stellar metallicity (\citealt{Bensby2017}). Its production occurs primarily in massive stars (\citealt{Bastian2018}), partly during hydrostatic carbon burning and partly in core-collapse supernovae (Type~II SNe; e.g. \citealt{Cameron1959, Salpeter1952, Woosley1995}).

The evolution of sodium abundance, expressed as [Na/Fe]\footnote{We adopt the customary elemental abundance notation where $A\mathrm{(X)} \equiv \log (N_\mathrm{X}/N_\mathrm{H}) + 12$ and $\mathrm{[X/Y]} \equiv (A(\mathrm{X}) - A(\mathrm{Y})) - (A(\mathrm{X}) - A(\mathrm{Y}))_\odot$, with $N_\mathrm{X}$ being the number density of element ``X''.} displays a number of features that reflect its diverse cosmic origins. At low metallicities ($-4 \lesssim \mathrm{[Fe/H]} \lesssim -1$), [Na/Fe] initially rises with increasing [Fe/H] (\citealt{Kobayashi2020}). This behaviour reflects the contributions of massive stars that synthesise both Fe and Na. The upward trend has been defined as the ``metallicity effect'' in \citet{Owusu2024}, wherein sodium production becomes more efficient in metal-rich environments. The only stable isotope, $^{23}\mathrm{Na}$, is neutron-rich (11 protons and 12 neutrons), so its synthesis is more efficient when additional neutrons are available.
At low-metallicity, during core He-burning, mainly self-conjugate nuclei like $^{12}\mathrm{C}$,$^{16}\mathrm{O}$, and $^{20}\mathrm{Ne}$ are produced. At higher [Fe/H], these C, N, and O nuclei are already present in the stellar core from the beginning, and therefore, during the core H-burning, the CNO cycle produces a significant build-up of $^{14}\mathrm{N}$.
Then, core He-burning turns $^{14}\mathrm{N}$ into the neutron-rich isotope $^{22}\mathrm{Ne}$, which contains two more neutrons than protons. This neutron-rich nucleus then serves as a more efficient target for proton capture, leading to the synthesis of $^{23}\mathrm{Na}$ via the reaction $^{22}\mathrm{Ne}(p, \gamma)^{23}\mathrm{Na}$.

After the initial rise of the [Na/Fe] ratio, it starts to decline near [Fe/H]~$\approx -1$, when Type~Ia SNe, which produce negligible amounts of Na, begin to contribute significantly to the interstellar enrichment (\citealt{Ruiter2011}). 
Finally, at approximately solar metallicity ([Fe/H]~$\approx 0$), when the relative rates of Type~II and Type~Ia SNe approach equilibrium, [Na/Fe] increases again (\citealt{Bensby2017}). This characteristic ``zig-zag'' abundance trend in the Milky Way (\citealt{McWilliam2016}) can be largely explained by the combined contributions of supernovae and hypernovae, with a minor enrichment component from super-asymptotic giant branch (AGB) stars (\citealt{Kobayashi2020}), with a metallicity-dependent yield.

Unique cases of sodium enrichment are observed in globular clusters (GCs) in the Galactic halo, where red giant branch (RGB) stars display star-to-star abundance variations indicative of multiple stellar populations. These variations suggest that distinct chemical enrichment processes have operated within such dense stellar environments (e.g. \citealt{Gratton2004}). Observations reveal at least two chemically distinct stellar populations in most GCs: a first population (1P) with compositions comparable to field stars, and a second population (2P) enriched in Na and N but depleted in C and O (e.g. \citealt{Gratton2001, Carretta2009, Gratton2011}). For a comprehensive review of multiple populations in star clusters, see \citet{Milone2022}.

The sodium enrichment observed in 2P stars is thought to originate from proton-capture nucleosynthesis through the Ne–Na cycle, which operated during high-temperature hydrogen burning in a previous stellar generation that mixed the processed material to the surface (e.g. \citealt{Denisenkov1990, ElEid1995, Mowlavi1999, Karakas2010}). However, the nature of the polluting sources responsible for this enrichment remains uncertain, with proposed candidates including AGB stars (e.g. \citealt{Bastian2018}), fast-rotating massive stars (e.g. \citealt{Decressin2007}), super-AGBs (e.g. \citealt{Pumo2008}), or massive interacting binaries (e.g. \citealt{deMink2009}). 

Accurate sodium abundances can be used to study these diverse production sites and shed light on stellar and Galactic physics. However, the accuracy of abundance determinations is sensitive to the underlying physical assumptions adopted in the modelling of radiative transfer and stellar atmospheres. A commonly used approximation is the local thermodynamic equilibrium (LTE), wherein the atomic level populations are determined by local conditions, following the Boltzmann and Saha equations. In stellar photospheres, however, radiative rates often dominate over collisional processes, causing departures from LTE. In such cases, the populations of atomic levels must instead be obtained by solving the statistical equilibrium equations, leading to non-LTE (NLTE) line formation (e.g. \citealt{Rutten2003}). Modelling line formation in NLTE is substantially more demanding, as the statistical equilibrium equations require iterative convergence of the level populations, and, additionally, accurate atomic and molecular data are required. Furthermore, classical one-dimensional (1D) hydrostatic model atmospheres, typically assuming plane-parallel or spherically symmetric geometry and treating convection through simplified prescriptions such as mixing-length theory (e.g. \citealt{BohmVitense1958}), remain widely used today. The use of such models also imparts errors into synthetic spectra, that can only be partially accounted for by calibrating the microturbulence and macroturbulence fudge parameters.
Thanks to code development and advances in computational resources over the last decades, it is now feasible to compute 3D radiation-hydrodynamic (RHD) simulations (e.g. \citealt{Magic2013}), which self-consistently model convective motions, temperature inhomogeneities, and velocity fields. Post-processing such models with 3D NLTE radiative transfer (e.g. \citealt{Lind2024}) naturally reproduces spectral-line broadening and asymmetries without the need for ad hoc parameters as well as more reliable equivalent widths and thereby more reliable abundance determinations that can shed light on stellar and Galactic physics (e.g. \citealt{Matsuno2024}).

Sodium is a minority species in late-type stellar atmospheres, and its spectral lines are particularly susceptible to NLTE effects, especially the strong \ion{Na}{I} D resonance lines (e.g. \citealt{Gratton1999, Mashonkina2000, Lind2011}). In 1D~LTE, these lines can yield abundance underestimates of up to 0.9\,dex in metal-poor ([Fe/H]~$=-2$) dwarfs and giants (\citealt{Marino2011}). 
The first full 3D~NLTE synthesis of \ion{Na}{I} lines was presented by \citet{Lind2013} for halo stars, revealing significant deviations from LTE. Subsequent 3D~NLTE studies of \ion{Na}{I} were performed by \citet{Nordlander2017} and \citet{Lagae2023} in two ultra-metal poor halo stars, where they help to constrain the properties of their Population III progenitors.
Recently, \citet{Asplund2021} carried out a full 3D~NLTE synthesis of \ion{Na}{I} for the Sun, leading to a slight downwards revision of the solar sodium abundance to $A(\mathrm{Na})_\odot = 6.22 \pm 0.03$\,dex compared to 6.24 in \citet{Asplund2009}. The centre-to-limb variation (CLV) of several solar \ion{Na}{I} lines was further investigated by \citet{Canocchi2024a}, who found excellent agreement between 3D~NLTE synthetic spectra and high-resolution, spatially resolved observations obtained with the Swedish 1-m Solar Telescope (SST; \citealt{Scharmer2003}). In contrast, LTE models—whether in 1D or 3D—significantly underestimate the strength of the \ion{Na}{I} D lines observed near the solar limb. These lines are highly sensitive to the atmospheric velocity fields, and thus, the traditional 1D plane-parallel models fail to reproduce their CLV even when NLTE is included.

To correct for the above-mentioned effects, several 1D~NLTE grids for \ion{Na}{I} have been published in the literature (e.g. \citealt{Mashonkina2000, Takeda2003, Lind2011, Lind2022}). The first grid of 3D~NLTE synthetic spectra, however, was recently computed by \citet{Canocchi2024b}, covering dwarfs ($4.0 < \log g < 5.0$) over $-0.5 < \mathrm{[Fe/H]} < +0.5$. This grid was used to correct for CLV effects in transmission spectra of four giant exoplanets observed with the ESPRESSO spectrograph (\citealt{Pepe2021}) at the Very Large Telescope, improving the precision of measured sodium abundances in three systems. In particular, the study demonstrated that 3D~NLTE stellar models alone can explain the sodium features in the transmission spectrum of HD~209458b without the need for additional planetary absorption. 
The same grid was later applied by \citet{Carlos2025} to correct sodium abundances for 50 F- and G-type stars with and without giant planets, revealing that the presence of giant planets exerts only a second-order effect compared to Galactic chemical evolution in shaping the observed abundance trends between planet-hosting stars and non-hosts.

In this work, we present an extended grid of 3D~NLTE synthetic spectra for several \ion{Na}{I} lines, computed using 3D RHD model atmospheres for FGK-type dwarfs and giants. This grid expands upon that of \citet{Canocchi2024b}, including metal-poor stars down to [Fe/H]$=-4$.
The paper is organized as follows. Section~\ref{sec: synthspectra} describes the model atmospheres, the model atom, and the radiative transfer code used for NLTE synthesis. Section~\ref{sec: interpolation} details the interpolation methods for synthetic spectra and equivalent widths. The results are presented in Section~\ref{sec: results}, and our conclusions are summarised in Section~\ref{sec: conclusions}.


\section{Synthetic stellar spectra}\label{sec: synthspectra}

\begin{figure}
	\includegraphics[width=1.1\columnwidth]{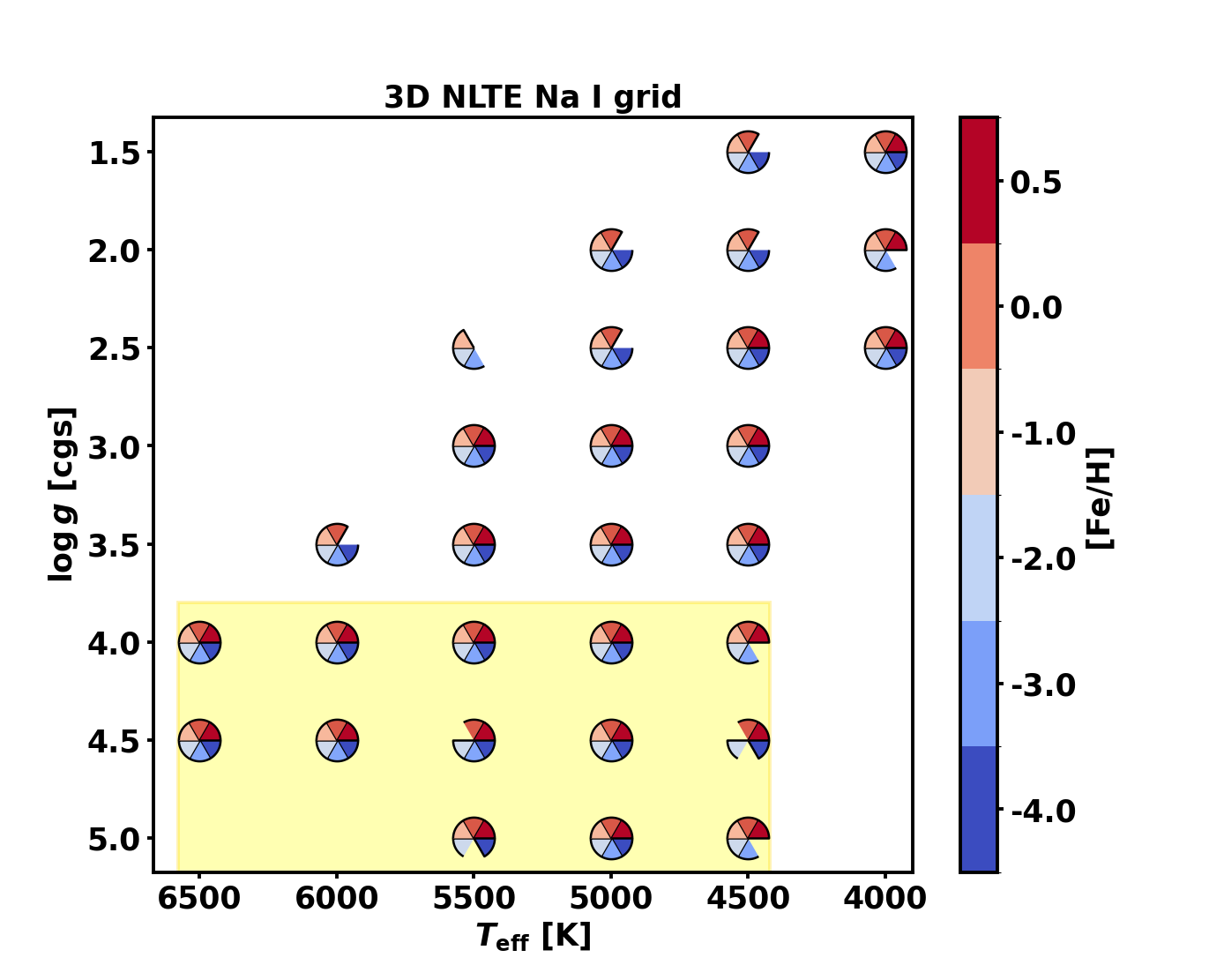}
    \caption{Kiel diagram illustrating the \texttt{Stagger}-grid nodes at which the 3D~NLTE computations have been performed, for values of [Fe/H] described by the colorbar. The models at [Fe/H]~$=0.0$ and $+0.5$ in the yellow-shaded area were previously published in \citet{Canocchi2024b}.}
    \label{fig: HRdiagram}
\end{figure}
\begin{figure*}
	\includegraphics[width=1.0\linewidth]{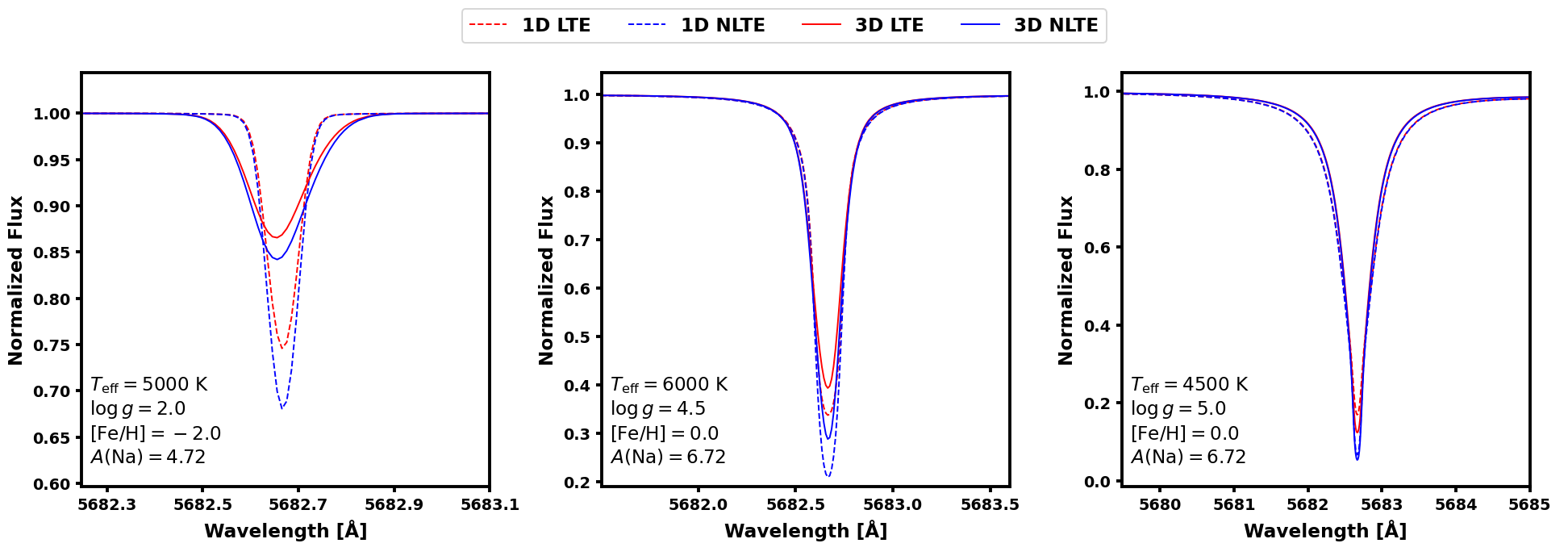} 
    \caption{Synthetic line profiles for the \ion{Na}{I} line at 5682\,\r{A}, computed with \texttt{Balder} in 1D LTE (dashed red line), 1D~NLTE (dashed blue line), 3D LTE (solid red line), and 3D~NLTE (solid blue line)  for a cool metal-poor red giant star (left), a solar metallicity F-dwarf (middle), and a metal-rich lower main-sequence dwarf (right). 1D $v_\mathrm{mic}=1.0$\,km\,s$^{-1}$, and a sodium abundance corresponding to [Na/Fe]$=+0.5$ are adopted. No rotational or macroturbulent broadening has been applied. Note that the y-axis has a different scale in each panel.} 
    \label{fig: line1balder}
\end{figure*}
\begin{table*}
\centering 
\caption{Atomic data of the sodium lines computed for the 3D~NLTE grid, as implemented in the model atom of \citet{Lind2011} with modifications from \citet{Canocchi2024a}.} 
\label{tab: lineproperties}
\begin{tabular}{l c c c c c c c} \hline \hline
\noalign{\smallskip} 
 $\lambda$ [\r{A}] & Transition & $^{(a)} E_\mathrm{low}$ [cm$^{-1}$] & $^{(b)}\log{gf}$ & $^{(c)} \sigma$ [a.u.] & $^{(d)} \alpha$ & $^{(e)} \Gamma$ [rad s$^{-1}$] & $^{(f)} \log{C_4}$ [cm$^4$ s$^{-1}$]\\ \hline
 \noalign{\smallskip} 
 5682 & $\mathrm{3p ^2P^0_{1/2}-4d^2D_{5/2}}$  & 16\,956.17 & $-0.706$ & 1955 & 0.327 & $8.05 \times 10^7$ & $-12.18$\\
 5688 & $\mathrm{3p  ^2P^0_{3/2}-4d^2D_{5/2}}$ & 16\,973.37 & $-0.452$ & 1955 & 0.327 & $8.07 \times 10^7$ & $-12.18$\\
 5889 & $\mathrm{3s  ^2S_{1/2}-3p^2P^0_{3/2}}$ & 0.00 & 0.108 & 407 & 0.273 & $6.16 \times 10^7$ & $-15.11$\\
 5896 & $\mathrm{3s  ^2S_{1/2}-3p^2P^0_{1/2}}$ & 0.00 & $-0.194$ & 407 & 0.273 & $6.14 \times 10^7$ & $-15.11$\\ 
 6154 & $\mathrm{3p  ^2P^0_{1/2}-5s^2S_{1/2}}$ & 16\,956.17 & $-1.547$ & 2594 & 0.021 & $7.43 \times 10^7$ & $-13.20$ \\
 6160 & $\mathrm{3p  ^2P^0_{3/2}-5s^2S_{1/2}}$ & 16\,973.37 & $-1.246$ & 2594 & 0.021 & $7.45 \times 10^7$ & $-13.20$ \\ 
 8183 & $\mathrm{3p^2P^0_{1/2} - 3d^2D_{3/2}}$ & 16\,956.17 &  0.237 &  804 & 0.270 & $1.13 \times 10^8$ & $-13.76$ \\
 8194 &$\mathrm{3p^2P^0_{3/2} - 3d^2D_{5/2}}$ & 16\,973.37 & 0.492 & 804 & 0.270 & $1.13 \times 10^8$ & $-13.76$\\ 
10\,747 & $\mathrm{4s ^2 S_{1/2}- 5p ^2P^0_{3/2}}$ & 25\,739.99 & $-1.294$ & 2720 & 0.192 & $2.91 \times 10^7$ & $-12.64$\\
\hline 
\end{tabular}\\
\begin{flushleft}
\textbf{Notes:}\\
$^{(a)} E_\mathrm{low}$ is the excitation energy of the lower level of the transition, taken from the NIST database (\citealt{Ralchenko2005, Sansonetti2008}). $^{(b)} \log{gf}$ are from the VALD3 database (\citealt{Ryabchikova2015}), where $f$ is the oscillator strength, or transition probability, and $g$ is the statistical weight. $^{(c)} \sigma$ is the broadening cross-section for elastic collisions with hydrogen atoms at a relative velocity of $10^4$ m s$^{-1}$. $^{(d)} \alpha$ is the exponent with which $\sigma$ varies with velocity (i.e., $v^{-\alpha}$, \citealt{Anstee1995}). $^{(e)} \Gamma$ is the natural broadening width. $^{(f)} C_4$ is the Stark broadening constant, estimated from \citet{Dimitrijevic1990}.

\end{flushleft}
\end{table*}

\subsection{3D hydrodynamical model atmospheres}\label{sec: 3Datmos}
The grid of synthetic spectra for \ion{Na}{I} presented in this work extends the 3D~NLTE grid of \citet{Canocchi2024b}. We employ 3D RHD model atmospheres from the publicly available\footnote{\url{https://3dsim.oca.eu/fr/the-stagger-grid-2-0}} extended \texttt{Stagger} grid (\citealt{Rodriguez2024}), computed with the \texttt{Stagger} code (\citealt{Galsgaard1995, Collet2011, Magic2013, Stein2024}) in a ``box-in-a-star'' setup. The models are characterised by their stellar parameters $T_\mathrm{eff}$, $\log g$, and [Fe/H], which serve as proxies for the overall chemical composition of the star. The adopted chemical composition follows the solar abundances of \citet{Asplund2009}, scaled by [Fe/H]; models with $\mathrm{[Fe/H]} \leq -1.0$ are $\alpha$-enhanced by $\mathrm{[\alpha/Fe]} = +0.4$. An overview of the grid used in this work is shown in Fig.~\ref{fig: HRdiagram}.
 
It is worth noting that in \texttt{Stagger} simulations the effective temperature is not an input parameter but an emergent property of the radiative-hydrodynamic solution. Consequently, the grid is not perfectly regular in $T_\mathrm{eff}$.

Each 3D model atmosphere is defined on a Cartesian grid of $240^3$ cells, encompassing at least ten convective granules and spanning several convective turnover times. For each model, five representative temporal snapshots were selected for the detailed 3D~NLTE radiative transfer calculations, following the procedure described in Sect.~4.4 of \citet{Rodriguez2024}.

To reduce computational costs in the spectrum synthesis, we downsampled the horizontal resolution of each snapshot from 240 to 48 cells, while increasing the vertical sampling by trimming away the optically-thick layers and interpolating onto a new depth scale of 240 vertical cells, with finer resolution of the steep continuum-forming regions. This set-up allows us to retain the refined resolution of the line-forming region (i.e. photosphere) where steep temperature gradients are present. The total number of volume elements is of the order of $5 \cdot 10^5$.
We also tested a higher horizontal resolution ($80 \times 80 \times 240$) and found differences of less than 1\% in the resulting line strengths and profiles compared to the $48 \times 48 \times 240$ models, consistent with the findings of \citet{Lagae2025}.
The computation of the entire grid of 3D~NLTE synthetic spectra took over 3 million CPU hours.

\subsection{1D hydrostatic model atmospheres}\label{sec: 1Datmos}
For completeness and comparison, we also employ 1D hydrostatic model atmospheres from the \texttt{MARCS} grid (\citealt{Gustafsson2008}). 
The same grid configuration was adopted by \citet{Amarsi2020} and \citet{Buder2021} in their 1D~NLTE analysis of the GALactic Archaelogy with HERMES (GALAH\footnote{\url{https://www.galah-survey.org/dr4/overview/}}; \citealt{DeSilva2015}) DR3 stars.
The \texttt{MARCS} models span the parameter ranges $3000 \leq T_\mathrm{eff}/\mathrm{K} \leq 8000$, $-0.5 \leq \log g/\mathrm{(cm,s^{-2})} \leq 5.5$, and $-5.0 \leq \mathrm{[Fe/H]} \leq +1.0$.
In 3D radiation-hydrodynamic models, convection arises naturally from first principles. In contrast, canonical 1D hydrostatic models approximate the effects of convection using the mixing-length formalism, and include empirical broadening parameters such as microturbulence ($v_\mathrm{mic}$). 
For dwarfs ($\log g \geq 4.0$), plane-parallel models with $v_\mathrm{mic} = 1$\,km\,s$^{-1}$ were adopted, while for giants ($\log g \leq 3.5$), spherically symmetric models with $v_\mathrm{mic} = 2$\,km\,s$^{-1}$ were used.

All models assume the solar chemical composition of \citet{Grevesse2007}, scaled by [Fe/H]. In agreement with the 3D grid, models with $\mathrm{[Fe/H]} \leq -1.0$ are $\alpha$-enhanced by $\mathrm{[\alpha/Fe]} = +0.4$. Each \texttt{MARCS} model is defined on a 1D depth scale of 56 layers, covering an optical depth range from $\log\tau_\mathrm{Ross} = -5$ to $+2$. Unlike the 3D RHD models, the 1D grids are regularly spaced in $T_\mathrm{eff}$. 

To enable a consistent comparison between the 1D and 3D results (see Sect.~\ref{sec: results}), the equivalent widths from the regularly spaced 1D grid were interpolated onto the irregular $T_\mathrm{eff}$–$\log g$–[Fe/H] points of the 3D grid using fully connected feed-forward neural networks (Sect.~\ref{sec: FFNN}).

\subsection{Model atom}\label{sec: modelatom}
A model atom is required for NLTE spectral synthesis. This model atom includes accurate atomic data for the energy levels and for the radiative and collisional transitions between them. We use the model atom for sodium developed in \citet{Lind2011}, with modifications to the hydrogen collisions described in \citet{Canocchi2024a}. The model comprises 23 energy levels in total: 22 levels for neutral sodium (\ion{Na}{I}) and one continuum level representing the ground state of ionised sodium (\ion{Na}{II}). 
Radiative and collisional transitions between these levels are taken into account. For inelastic collisions with neutral hydrogen, we adopt the rate coefficients from \citet{Barklem2010}, which are based on the quantum-scattering cross-sections computed by \citet{Belyaev2010}. Collisional excitation and ionisation by electrons are treated using the data from \citet{Igenbergs2008} and \citet{Gao2010}.
More details about the collisional transitions can be found in Sect.~2.3 and 2.4 of \citet{Lind2011}. 
Moreover, the model atom also includes 166 bound-bound radiative transitions, with most of the oscillator strengths from the ab initio calculations of Froese Fischer\footnote{Multi-configuration Hartree-Fock computations (MCHF).  \url{http://www.vuse.vanderbilt.edu/~cff/mchf_collection/}}, and the remaining ones from TOPbase atomic database\footnote{\url{http://legacy.gsfc.nasa.gov/topbase}} (\citealt{Cunto1992}). The only exception is for the \ion{Na}{I} D lines at 5889\,\r{A} and 5896\,\r{A}, for which accurate experimental data are adopted from the NIST database\footnote{\url{http://physics.nist.gov/PhysRefData/ASD/index.html}}.
Table \ref{tab: lineproperties} summarizes the atomic properties for the nine main \ion{Na}{I} transitions considered in this work.

\subsection{Spectrum synthesis}\label{sec: synthspectra2}
For the spectrum synthesis, we make use of the 3D~NLTE radiative transfer MPI-parallelized code \texttt{Balder} (\citealt{Amarsi2018}), which is a stellar offshoot of \textsc{Multi3D} (\citealt{Botnen1999, Leenarts2009}) that is primarily used for studying the Solar chromosphere.  
The statistical equilibrium is solved by calculating the mean radiation field $J$ with short characteristic rays, using the eight-point Lobatto quadrature on the interval [$-1, 1$] for the integration over $\mu=\cos{\theta}$\footnote{$\theta$ is defined as the angle between the outward normal of the stellar surface and the observer’s line of sight.}, until convergence of the level population is reached. 
After the last iteration, long characteristic rays are used to compute the final emergent intensity spectra. Then, the astrophysical flux is computed by disc-integrating the emergent intensities. A seven-point Lobatto quadrature is employed over the interval [0, 1] for integration with respect to $\mu$, while an eight-point equidistant trapezoidal quadrature over the interval [0, $2\pi$] is used for integrating the azimuthal angle $\phi$ for non-vertical rays. For a more detailed overview of the code, we refer to Sect.~2.2 in \citet{Amarsi2018}.

We compute 3D~NLTE spectra for abundances [Na/Fe]= $-0.5$ to $+0.5$ for dwarfs ($\log g \geq 4.0$), and [Na/Fe]$=-0.5$ to 1.0 for giants ($\log g \leq 3.5$), in steps of 0.5\,dex.
The input abundance value in \texttt{Balder} is given in absolute abundance $A(\mathrm{Na})$, which varies from model to model in the grid, depending on the [Fe/H], as shown in detail in Table \ref{tab: synthspectraparams} in Appendix.
The same code, \texttt{Balder}, was employed to compute spectra in 3D~LTE, 1D~NLTE, and 1D~LTE.  
Figure~\ref{fig: line1balder} compares the resulting line profiles at a fixed abundance for the \ion{Na}{I} 5682\,\r{A} line in three representative cases: weak, saturated, and strong lines.

In the left panel, showing a metal-poor giant, the line lies in the linear part of the curve of growth. Here, the abundance correction between 1D~LTE and 3D~NLTE is small (less than $-0.1$\,dex; see Sect.~\ref{sec: 3DNLTEcorr}). Without macroturbulence applied to the 1D models, the natural broadening and asymmetry of the 3D lines, directly arising from the hydrodynamical simulation, are clearly visible. 
In particular, the 3D line profiles appear slightly blueshifted relative to their 1D counterparts, owing to convective granulation at the stellar surface. The bright, rising hot gas in the granules contributes more to the emergent flux than the sinking, cooler gas in the intergranular lanes. This imbalance leads to a net upward motion in the spatially unresolved stellar disk, and hence to an overall convective blueshift (\citealt{Dravins1981, Dravins2021}).

The middle panel presents a solar-metallicity dwarf with saturated lines. In this case, the main differences appear in the line core: NLTE models yield a much deeper core. 
As in the previous case, NLTE produces stronger lines than LTE, leading to negative abundance corrections relative to 1D LTE.
Finally, the right panel illustrates a metal-rich dwarf with strong lines in the damping part of the curve of growth. Here, the wings of the 1D profiles are stronger than those of the 3D models. Although the 3D~NLTE core is deeper (i.e. formed in higher layers), the overall equivalent width of the 1D LTE line is larger, resulting in positive abundance corrections.

\section{Interpolation}\label{sec: interpolation}
\begin{figure*}
    \centering
    \includegraphics[width=1.0\linewidth]{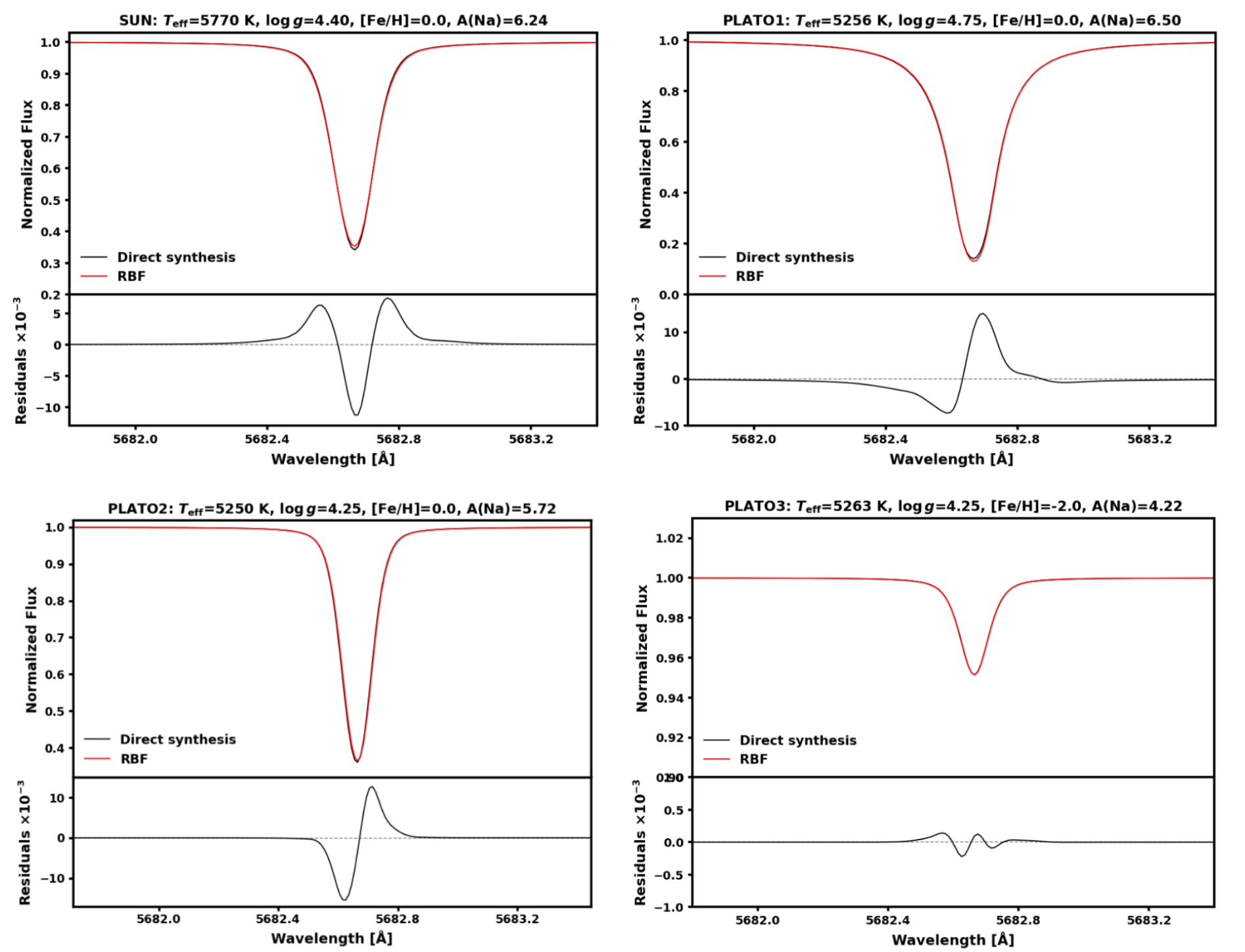}
    \caption{Comparison of interpolation models with verification models (not included in the training set) for the \ion{Na}{I} line at 5682\,\r{A}. The verification models were tailored to specific stars, indicated by the stellar parameters: $T_\mathrm{eff}, \log g,$ [Fe/H], and $A\mathrm{(Na)}$. Note that the range of the y-axis differs between panels.}
    \label{fig: verification1}
\end{figure*}
\begin{table} 
    \centering
    \caption{Error statistics of the RBF in predicting sodium abundance ($A\mathrm{(Na)}$) from the predicted normalized flux in the leave-one-out cross-validation on the 3D~NLTE grid. The columns show the root-mean-square (RMS) and the median absolute abundance deviation (MAD) of the corresponding RBF model with optimized hyperparameter $s$.} \label{tab: RBFl1o}
    \begin{tabular}{l c c c c} \hline \hline
    \noalign{\smallskip}
    $\lambda$ & RMS(${\Delta A\mathrm{(Na)}})$ & MAD$({\Delta A\mathrm{(Na)}}$) & $s$\\  
    {[}\AA{]} & [dex] & [dex] & \\ \hline
    \noalign{\smallskip} 
     5682  & 0.060 & 0.012 & $10^{-5}$ \\ 
     5688  & 0.059 & 0.013 & $10^{-5}$ \\ 
     5889 & 0.061 & 0.027 & $10^{-1}$ \\
     5896 & 0.059 & 0.026 & $10^{-1}$ \\ 
     6154 & 0.051 & 0.012 & $10^{-8}$\\ 
     6160 & 0.054 & 0.013 & $10^{-7}$ \\ 
     8183 & 0.083 & 0.015 & $10^{-4}$ \\
     8194 & 0.084 & 0.017 & $10^{-3}$ \\ 
     10\,747 & 0.037 & 0.009 & $10^{-10}$ \\ \hline  
     \hline
    \end{tabular}
\end{table}
\begin{table} 
    \centering
    \caption{Error statistics of the FFNN in predicting abundance ($A\text{(Na)}$) in the leave-one-out cross-validation on the 3D~NLTE grid. The columns show the root-mean-square (RMS) and the median absolute deviation (MAD) of the corresponding FFNN with optimized hyperparameters (number of layers, $n_l$, neurons per layer, $n$, and L2 penalty, $\alpha$).}
    \label{tab: FFNNl1o}
    \begin{tabular}{l c c c c c c} \hline \hline
    \noalign{\smallskip}
    $\lambda$ & RMS(${\Delta A\mathrm{(Na)}}$) & MAD(${\Delta A\mathrm{(Na)}}$) & $n_l$ & $n$ & $\alpha$ \\ 
    {[}\AA{]} & [dex] & [dex] & & & \\ \hline 
    \noalign{\smallskip} 
     5682  & 0.027 & 0.008 & 3 & 400 & 0.001 \\ 
     5688 & 0.028 & 0.007 & 2 & 400 & 0.001 \\ 
     5889 & 0.032 & 0.016 & 2 & 500 & 0.01 \\ 
     5896 & 0.036 & 0.015 & 3 & 200 & 0.0001 \\ 
     6154 & 0.023 & 0.008 & 2 & 500 & 0.001 \\ 
     6160 & 0.026 & 0.008 & 3 & 400 & 0.001 \\ 
     8183 & 0.038 & 0.009 & 3 & 200 & 0.0001\\ 
     8194 & 0.037 & 0.011 & 3 & 100 & 0.001\\ 
     10\,747 & 0.020 & 0.005 & 2 & 500 & 0.0001\\ \hline  
    \hline
    \end{tabular}
\end{table}
\begin{table*}
\centering 
\caption{Error in the predicted sodium abundance ($A(\mathrm{Na})$) obtained using the RBF and FFNN interpolation models for verification stars excluded from the training set. The first and second rows (‘Direct synthesis’) display the input abundances, while subsequent rows show the deviation from the reference values ($\Delta A\mathrm{(Na)}$). The stellar parameters of the verification stars are presented in Fig.~\ref{fig: verification1}.}\label{tab: verification}
\begin{tabular}{l c c c c c} \hline \hline
  \noalign{\smallskip}
 Model & & Sun & PLATO1 & PLATO2 & PLATO3\\ 
 $\lambda$[\r{A}] & Method  & & & & \\ \hline
 \noalign{\smallskip}  
  & Direct synthesis & 6.24 & 6.50 & 5.72 & 4.22 \\ 
  & & & & & \\
  5682 & RBF & 0.025 & $-0.003$ & 0.007 & 0.003 \\ 
   & FFNN & 0.003 & $-0.004$ & 0.016 & $-0.006$ \\ 
  5688 & RBF & 0.026 & 0.002 & 0.007 & 0.005 \\ 
   & FFNN & $-0.007$ & 0.001 & 0.012 & 0.002 \\ 
  5889 & RBF & 0.014 & 0.023 & 0.004 & 0.010 \\ 
  & FFNN & 0.013 & 0.019 & 0.010 & 0.003 \\ 
  5896 & RBF & 0.014 & 0.019 & 0.005 & 0.009 \\ 
  & FFNN & 0.014 & $-0.016$ & 0.011 & 0.009 \\ 
  6154 & RBF & 0.028 & 0.003 & 0.001 & 0.004\\ 
  & FFNN & $-0.020$ & 0.004 & 0.002 & $-0.003$ \\ 
  6160 & RBF & 0.019 & $-0.006$ & 0.003 & 0.006 \\ 
  & FFNN & $-0.014$ & 0.010 & 0.002 & 0.001 \\ 
  8183 & RBF & 0.030 & 0.004 & 0.007 & 0.014 \\
  & FFNN & 0.009 & 0.006 & 0.017 & 0.008\\
  8194 & RBF & 0.034 & 0.008 & 0.007 & 0.023 \\ 
  & FFNN & $-0.017$ & 0.008 & 0.015 & 0.001 \\ 
  10\,747 & RBF & 0.022 & 0.009 & 0.001 & 0.001 \\ 
  & FFNN & $-0.015$ & 0.001 & 0.001 & $-0.009$ \\ \hline  
  \hline
\end{tabular}
\end{table*} 

The \texttt{Stagger} grid is tabulated at irregular intervals in $T_\mathrm{eff}$, with slightly different effective temperatures for each grid node (i.e. for each combination of $\log g$ and [Fe/H]). This arises because, in the 3D RHD stellar atmospheres of the \texttt{Stagger} grid, the effective temperature is an output rather than an input parameter. Consequently, the spacing in $T_\mathrm{eff}$ is not exactly 500\,K, as illustrated in Fig.~\ref{fig: HRdiagram}. Moreover, the grid is relatively sparse in parameter space, making the interpolation for this grid more complicated. 
As shown by \citet{Wang2021}, simple methods such as linear or spline interpolation tend to yield a larger error in abundance on the edges of grids of synthetic spectra based on the \texttt{Stagger} grid, compared to more sophisticated interpolation methods. 

Following the results of \citet{Wang2021, Wang2024}, we adopt two distinct approaches for interpolating spectral line profiles and line strengths, respectively. The interpolation of spectral line profiles, which involve the extra dimension of wavelengths and exhibit non-smooth variations due to broadening processes and hydrodynamic velocity fields, is performed using radial basis functions (RBFs) following the implementation of \citet{deLis2022}.
In contrast, line strengths, quantified as reduced equivalent widths (REWs), are interpolated using fully connected feed-forward neural networks (FFNNs). REWs are defined as:
\begin{equation}
    \mathrm{REW} = \log_{10} \left ( \frac{W_{\lambda_0}}{\lambda_0} \right ) \ ,
    \label{eq: REW}
\end{equation}
where $W_{\lambda_0}$ is the equivalent width of the line profile and $\lambda_0$ its central wavelength at rest.

The hyperparameters for each method are optimized using standard procedures (e.g. \citealt{james2013}), including five-fold cross-validation with a fixed random seed for reproducibility. The average interpolation error is estimated via leave-one-out cross-validation. Final RBF and FFNN models are trained on all available nodes of the \texttt{Stagger} grid and subsequently validated against additional \texttt{Stagger} models tailored to specific stars that were not included in the training set. The selected verification models represent: the Sun, a solar-metallicity Na-rich dwarf (PLATO1), a solar-metallicity Na-poor dwarf (PLATO2), and a metal-poor dwarf (PLATO3). 
The three models labelled PLATO1, PLATO2, and PLATO3 take their names from the fact that they were newly added to the \texttt{Stagger} grid by \citet{Rodriguez2024} in order to refine the region of parameter space most relevant for the PLATO space mission (\citealt{PLATOref}).
The interpolation of line profiles and line strengths is discussed in more detail in Sects.~\ref{sec: RBF} and~\ref{sec: FFNN}, respectively.
The verification of the final RBF and FFNN models trained on the full grid is discussed in Sect.~\ref{sec: verification}.

\subsection{Interpolation of spectral line profiles}\label{sec: RBF}
Following the results of the interpolation tests performed in \citet{Wang2021}, in order to make the interpolation of the line profiles easier, we transform the normalized flux $f$ to a quantity ($f_t$: transformed flux) that scales better with abundance:
\begin{equation}
    f_t = \log_{10} (1 - f + s) \ ,
    \label{eq: transformflux}
\end{equation}
where $s$ is a small positive factor (treated as a hyperparameter) that smoothly truncates $f_t$ as the flux approaches the continuum. 
We tested interpolating over the normalized flux $f$, finding larger relative errors when the flux depression is small, thus turning into larger abundance errors as well as larger errors in the detailed shape for weak lines. 

RBF models were trained to interpolate transformed line profiles as functions of the stellar parameters ($T_\mathrm{eff}$, $\log g$, [Fe/H]) and sodium abundance ($A(\mathrm{Na})$). RBF model optimisation employed five-fold cross-validation across ten values of the $s$ parameter ($10^{-10} \leq s \leq 10^{-1}$). 
For each \ion{Na}{I} line, we adopt the optimal value of $s$ as the one with the lowest five-fold cross-validation median absolute deviation (MAD). 
The model accuracy was evaluated using the MAD and the root-mean-square error (RMS), with the MAD adopted for the selection of hyperparameters owing to its reduced sensitivity to outliers and better representation of the typical interpolation error.

Table~\ref{tab: RBFl1o} summarises the optimal $s$-values for each \ion{Na}{I} line and the error statistics of the RBF model, estimated using a leave-one-out cross-validation procedure. In this approach, one model is systematically excluded from the dataset, the interpolation is performed on the remaining models, and the line profiles corresponding to the omitted stellar parameters are predicted.

\subsection{Interpolation of line strengths}\label{sec: FFNN}
FFNN models were trained to predict the 3D~NLTE $A(\mathrm{Na})$ from a given set of stellar parameters ($T_\mathrm{eff}$, $\log g$, [Fe/H]) and reduced equivalent widths (REW). Multilayer perceptrons were implemented using the \textsc{MLPRegressor} class (\citealt{Hinton1990}) from the \textit{scikit-learn} package\footnote{\url{https://scikit-learn.org/stable/modules/generated/sklearn.neural_network.MLPRegressor.html}} (\citealt{Pedregosa2011}). The models were trained with a maximum of $10^5$ iterations, a convergence tolerance of $10^{-6}$, and employed the rectified linear unit (ReLU) as the activation function.

Similarly to the RBF analysis, a five-fold cross-validation was carried out to optimise the three main hyperparameters: the number of layers ($n_l$), the number of neurons per layer ($n$), and the L2 regularisation term ($\alpha$). The optimal configurations for each \ion{Na}{I} line, together with the corresponding leave-one-out cross-validation error statistics, are presented in Table~\ref{tab: FFNNl1o}. 

In addition, supplementary FFNN models were trained on the 1D~LTE, 1D~NLTE, and 3D~LTE grids to enable direct comparison with the 3D~NLTE abundances. The best hyperparameters for these models are presented in Table~\ref{tab: FFNNothermodels} in Appendix.
These models were used to compute abundance corrections for each \ion{Na}{I} line, defined as the abundance difference between two modelling assumptions—e.g. between 3D~NLTE and 1D~LTE—at fixed equivalent width,
\begin{equation}
\Delta^\mathrm{3N}_\mathrm{1L} = A(\mathrm{Na})_\mathrm{3DNLTE} - A(\mathrm{Na})_\mathrm{1DLTE} \ ,
\label{eq:corrAb}
\end{equation}
where $A(\mathrm{Na})$ is derived by matching the REW along the curve of growth between two models with identical stellar parameters ($T_\mathrm{eff}$, $\log g$, [Fe/H], and 1D $v_\mathrm{mic}$). 
The equivalent width of the \ion{Na}{I} synthetic lines was obtained by numerical integration over the line profile, considering a spectral region that extends $\pm 4.0$\,\r{A} from the line centre for the \ion{Na}{I} D lines, and $\pm 2.0$\,\r{A} for the other lines.

\subsection{Verification of final interpolation models}\label{sec: verification}

To further assess the validity of the final interpolation models and characterise the expected errors, we employed a verification set of four models from the recently extended \texttt{Stagger} grid, not included in the training set. Figure~\ref{fig: verification1} compares the exact line profiles synthesised with \texttt{Balder} for these verification models to the profiles predicted by the RBF interpolation, for the line at 5682\,\r{A}. The corresponding measured abundance errors are listed in Table~\ref{tab: verification}. In most cases, the errors for the verification models are smaller than the leave-one-out cross-validation errors, indicating that the final interpolation model yields consistent results when evaluated against independent stellar models. On average, the final RBF model predicts abundances in the verification sample with an error of approximately 0.013\,dex.

The final FFNN models trained on the full grid were validated using the same four verification stars as in the RBF analysis. The resulting abundance errors are listed in Table~\ref{tab: verification}. On average, the FFNN errors are smaller than those from the RBF models, and in most cases, the verification errors are lower than the leave-one-out cross-validation estimates, confirming that the final interpolation models retain high accuracy when applied to independent data.
In summary, the typical error in abundance determination for the FFNN is on average  $\sim$0.011\,dex at solar-metallicity, but less than 0.009\,dex in the metal-poor regime.


\section{Results and discussion}\label{sec: results}
\begin{figure*} 
    \centering
    \includegraphics[width=1.0\linewidth]{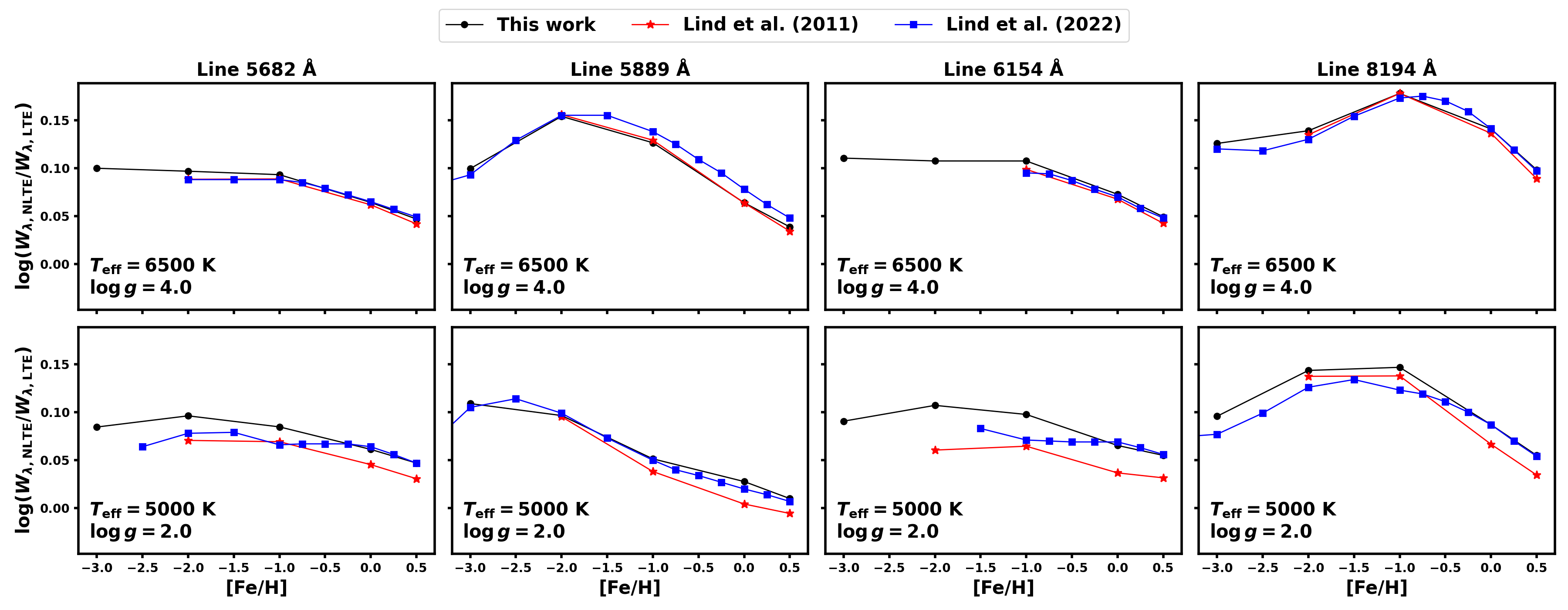}
    \caption{Comparison of the REWs in 1D~NLTE and 1D~LTE for four selected \ion{Na}{I} lines for a turn-off star (top) and a giant (bottom). The REWs are computed assuming [Na/Fe]$=0.0$ and $v_\mathrm{mic}=1.0$\,km\,s$^{-1}$. Black points show the REWs obtained in this work, while the red stars and blue squares denote results from \citet{Lind2011} and \citet{Lind2022}, respectively, as indicated in the legend.}
    \label{fig: comparison1D}
\end{figure*}
\begin{figure*} 
    \centering
    \includegraphics[width=0.9\linewidth]{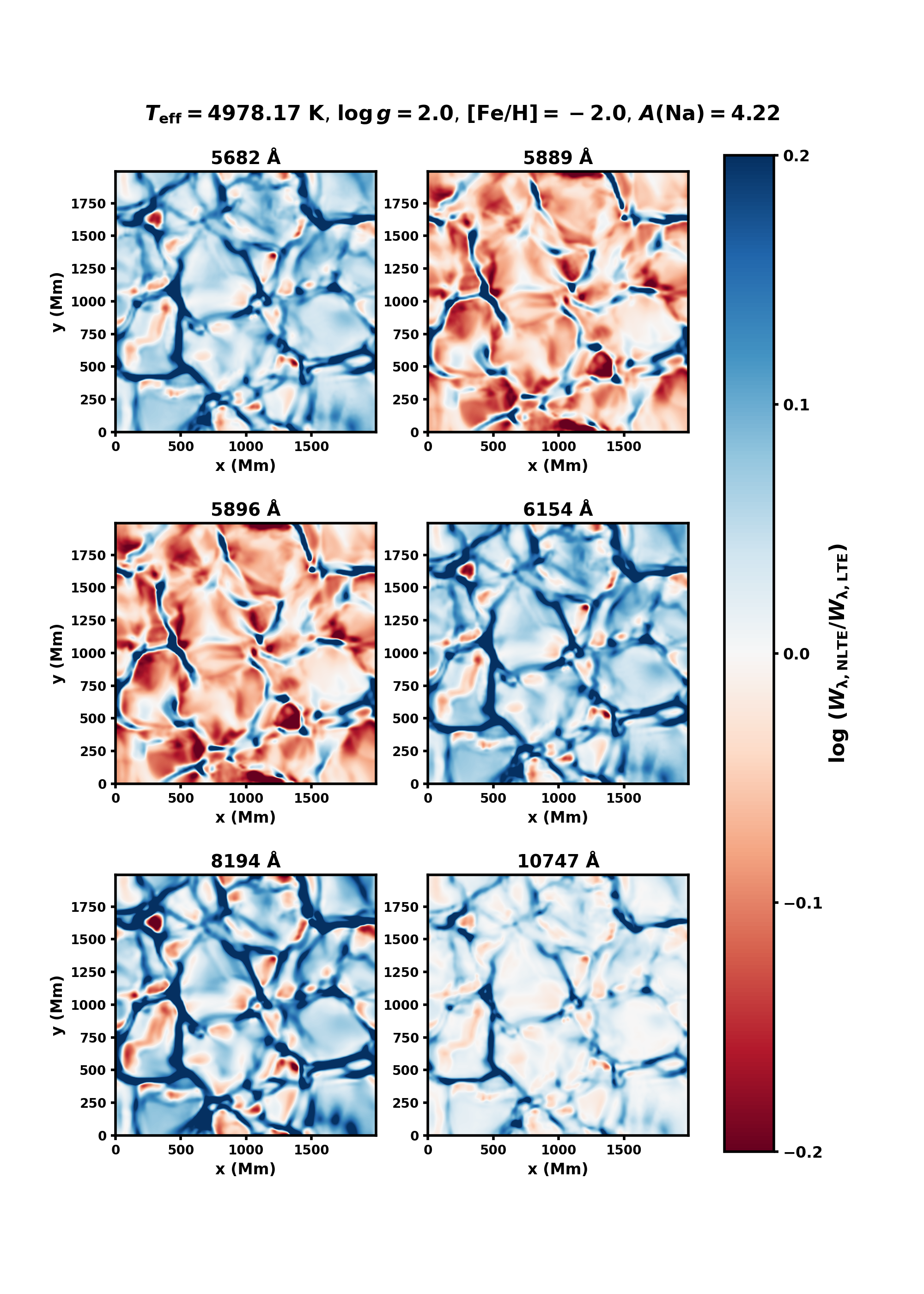}
    \caption{Spatially resolved difference between the reduced equivalent width in NLTE and LTE at disc centre intensity for different \ion{Na}{I} lines in a 3D model with $T_\mathrm{eff} = 4978.17$\,K, $\log g=2.0$, [Fe/H]$=-2.0$ and $A$(Na)=4.22.}
    \label{fig: resolvedNa}
\end{figure*}
\begin{figure} 
    \centering
    \includegraphics[width=1.0\linewidth]{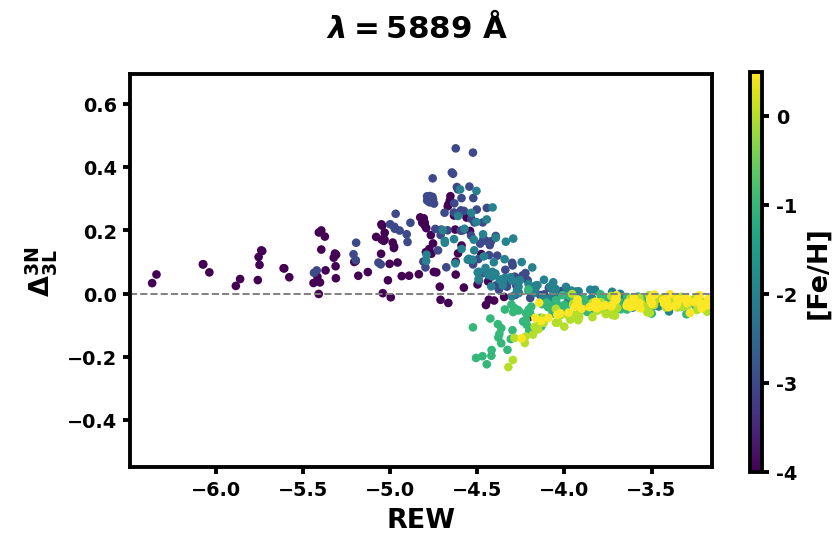}
    \caption{Abundance corrections for the \ion{Na}{I}~D line at 5889\,\r{A} for all models between 3D~NLTE to 3D~LTE, colour-coded by their [Fe/H].  The datapoints are computed at [Na/Fe]$=-0.5$ to $+0.5$ for the dwarfs, and $+1.0$ for the giants, in steps of 0.5\,dex.}
    \label{fig: 3N3Labcorr5889}
\end{figure}
\begin{figure*} 
    \centering
    \includegraphics[width=1.0\linewidth]{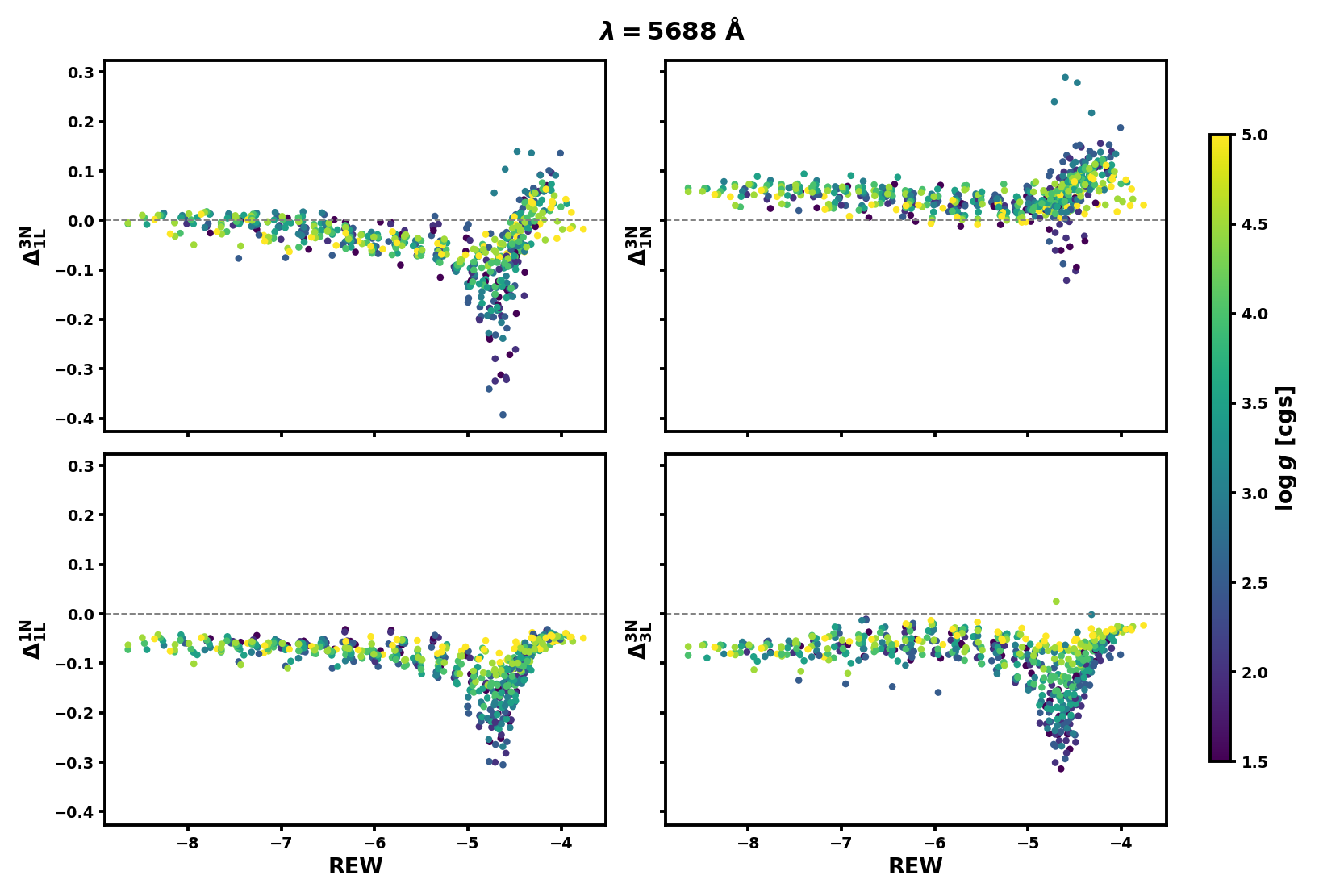}
    \caption{Abundance corrections for the \ion{Na}{I} line at 5688\,\r{A} for all models between: 3D~NLTE to 1D~LTE (top left), 3D~NLTE to 1D~NLTE (top right), 1D~NLTE to 1D~LTE (bottom left) and 3D~NLTE to 3D~LTE (bottom right), colour-coded by their $\log{g}$. The datapoints are computed at [Na/Fe]$=-0.5$ to $+0.5$ for the dwarfs, and $+1.0$ for the giants, in steps of 0.5\,dex and, for the 1D models, at a $v_\mathrm{mic}=1.0$\,km\,s$^{-1}$.}
    \label{fig: ALLabcorr5688}
\end{figure*}
\begin{figure*} 
    \centering
    \includegraphics[width=1.0\linewidth]{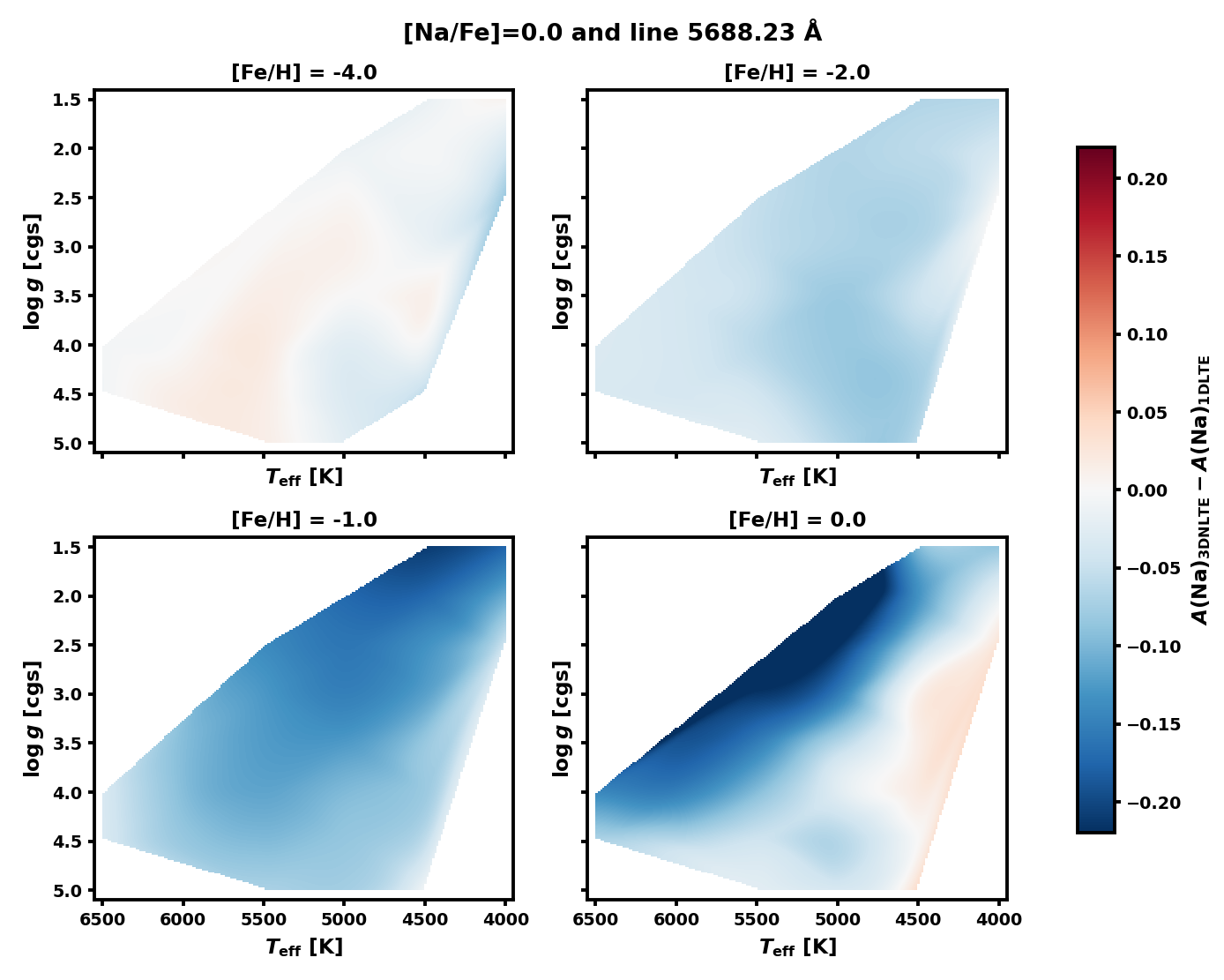}
    \caption{3D~NLTE - 1D~LTE abundance corrections ($\Delta^\mathrm{3N}_\mathrm{1L}$) for the 5688\,\r{A} line, shown at [Na/Fe]$=0.0$ and 1D models with $v_\mathrm{mic} = 1.0$\,km\,s$^{-1}$. Each panel corresponds to a different metallicity, and specifically, from left to right, top to bottom: [Fe/H]$= -4.0, -2.0, -1.0, 0.0$.} 
    \label{fig: abcorr5688}
\end{figure*}
\begin{figure} 
    \centering
    \includegraphics[width=1.0\linewidth]{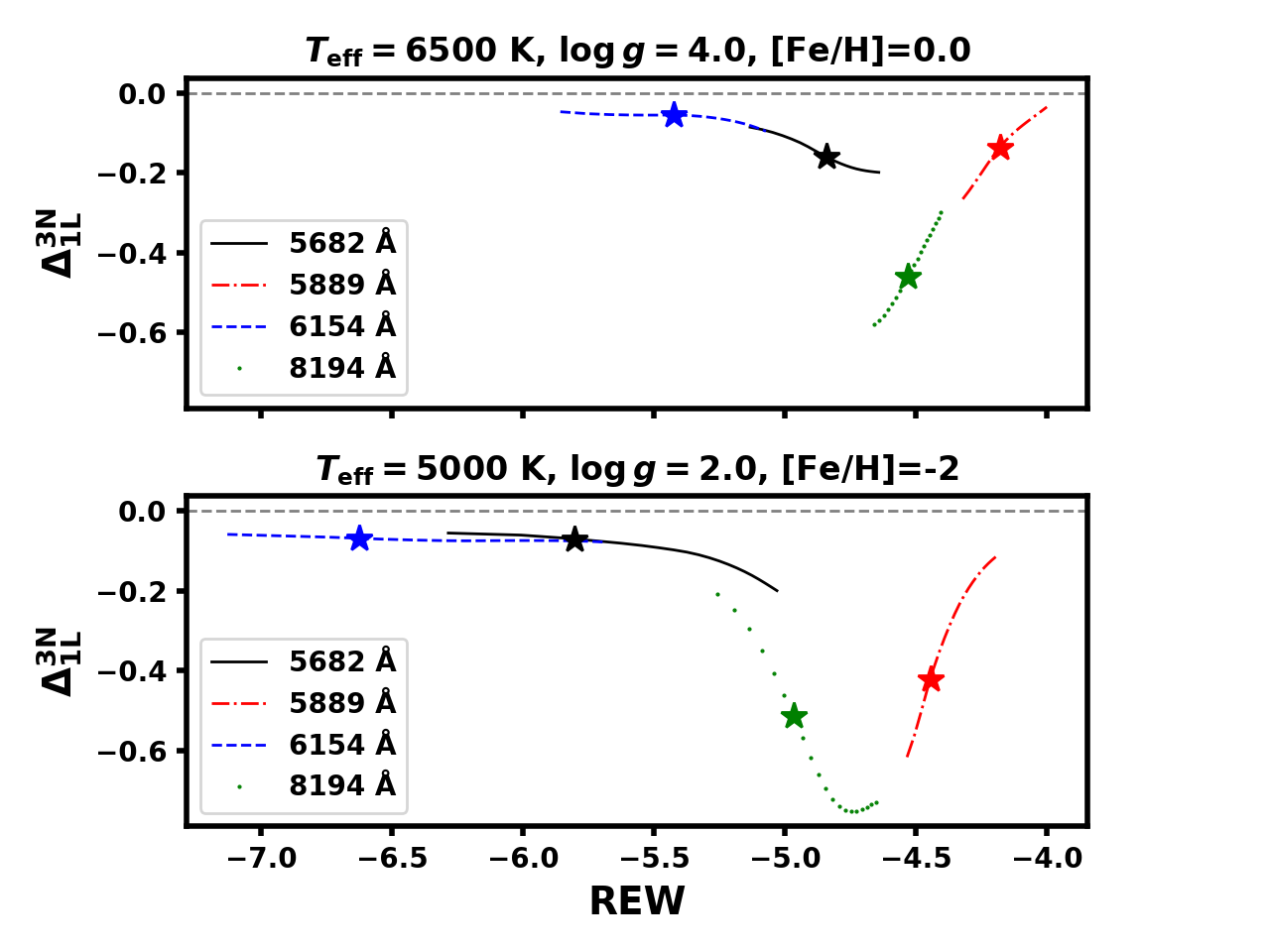}
    \caption{3D~NLTE abundance corrections as functions of reduced equivalent widths for four selected \ion{Na}{I} lines for a solar-metallicity turn-off star (top) and a metal-poor giant  (bottom). The star markers indicate the predicted line strength and abundance corrections at [Na/Fe] = 0.}
    \label{fig: 2metpoor}
\end{figure}
\begin{figure*} 
    \centering
    \includegraphics[width=1.0\linewidth]{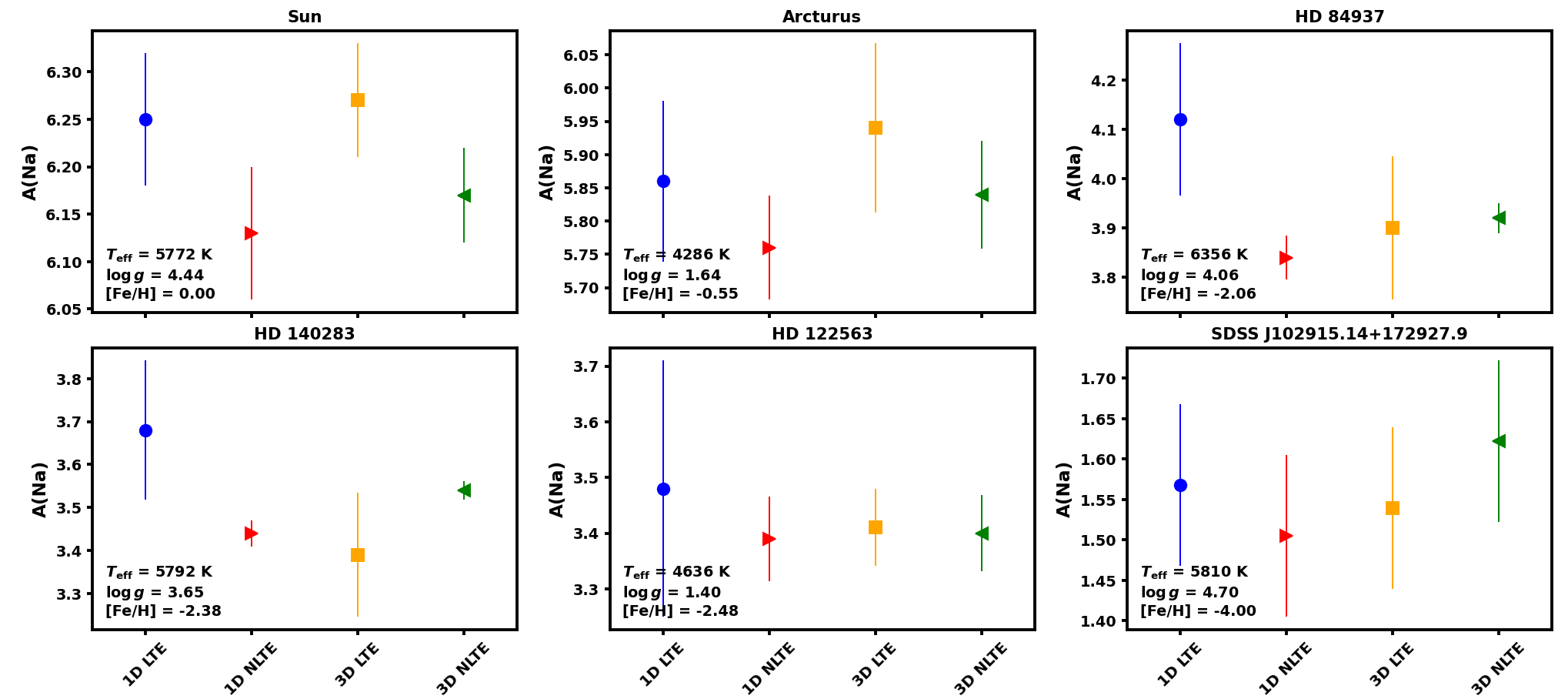}
    \caption{\ion{Na}{I} abundances in 1D~LTE, 1D~NLTE, 3D~LTE, and 3D~NLTE for our selected benchmark stars and the ultra metal-poor star SDSS J102915.14+172927.9. The derived $A\mathrm{(Na)}$ represents the weighted mean and the one sigma standard deviation of the available lines measured as in \citet{Lind2022}.} 
    \label{fig: benchmarks}
\end{figure*}
\begin{table*} 
    \centering
    \caption{Adopted stellar parameters and derived sodium abundances ($A\mathrm{(Na)}$) of the benchmark stars and the ultra metal-poor star analysed in this work.}\label{tab: benchmarks}
    \begin{tabular}{l c c c c c c c} \hline \hline
    \noalign{\smallskip}
    Star & $T_\mathrm{eff}$ & $\log g$ & [Fe/H] & \multicolumn{4}{c}{$A\mathrm{(Na)}$} \\  
     & [K] & [cm s$^{-2}$] & & 1D LTE & 1D NLTE & 3D LTE & 3D NLTE \\ \hline
    \noalign{\smallskip} 
     $^{(a, d)}$Sun  & 5772 & 4.44 & 0.0 & $6.25 \pm 0.07$ & $6.13 \pm 0.07$ & $6.27 \pm 0.06$ & $6.17 \pm 0.05$\\ 
     $^{(b,d)}$Arcturus & 4286 & 1.64 & $-0.55$ & $5.86 \pm 0.12$ & $5.76 \pm 0.08$ & $5.94 \pm 0.13$ & $5.84 \pm 0.08$\\ 
     $^{(b,d)}$HD~84937 & 6356 & 4.06 & $-2.06$ & $4.12 \pm 0.15$ & $3.84 \pm 0.05$ & $3.90 \pm 0.14$ & $3.92 \pm 0.03$\\ 
     $^{(c,d)}$HD~140283 & 5792 & 3.65 & $-2.38$ & $3.68 \pm 0.16$ & $3.44 \pm 0.03$ & $3.39 \pm 0.14$ & $3.54 \pm 0.02$\\ 
     $^{(c,d)}$HD~122563 & 4636 & 1.40 & $-2.48$ & $3.48 \pm 0.23$ & $3.39 \pm 0.08$ & $3.41 \pm 0.07$ & $3.40 \pm 0.07$\\  
     $^{(e)}$SDSS J102915.14+172927.9 & 5810 & 4.70 & $-4.00$ & $< 1.56 \pm 0.10$ & $< 1.50 \pm 0.10$ & $<1.54 \pm 0.10$ & $<1.62 \pm 0.10$\\
     \hline  
    \end{tabular}\\
    \begin{flushleft}
\textbf{Notes:}\\
$^{(a)}$ Solar $T_\mathrm{eff}$ and $\log g$ from \citet{Prsa2016}. $^{(b)}$ $T_\mathrm{eff}$ and $\log g$ from \citet{Heiter2015}. $^{(c)}$ $T_\mathrm{eff}$ and $\log g$ from \citet{Karovicova2020}. $^{(d)}$ [Fe/H] from \citet{Lind2022}. $^{(e)}$ Stellar parameters from \citet{Lagae2023}.
In the 1D models, $v_\mathrm{mic}=1.0$\,km\,s$^{-1}$ was adopted.
\end{flushleft}
\end{table*}
In this work, we present a new, state-of-the-art grid of 3D~NLTE synthetic spectra and abundances for \ion{Na}{I}. The effects of NLTE on line formation in 1D and 3D are investigated in Sects.~\ref{sec: 1Ddepartures} and \ref{sec: 3Ddepartures}, respectively. Abundance corrections and recommendations on suitable lines for sodium abundance analysis are provided in Sects.~\ref{sec: 3DNLTEcorr} and \ref{sec: recommendations}, respectively.
The validation of our 3D~NLTE models on a set of benchmark stars is performed in Sect.~\ref{sec: benchmark}.

In the following discussion, we define weak lines as having a REW smaller than $-5$, saturated lines when $-5 \leq \mathrm{REW} \leq-4$, and strong lines with REW higher than $-4$, meaning that they lie in the damping part of the curve of growth with the development of broad wings. 

\subsection{Departures from LTE in 1D}\label{sec: 1Ddepartures}
The effects of NLTE on \ion{Na}{I} lines in 1D models have been explored extensively in the literature, beginning over fifty years ago with the work of \citet{Johnson1964} and \citet{Athay1969} on the resonance lines. A few years later, \citet{Gehren1975} investigated departures from LTE for several \ion{Na}{I} lines in the solar atmosphere. Since then, numerous NLTE studies have been performed under the assumption of 1D stellar atmospheres, primarily for \ion{Na}{I} lines in the optical and near-infrared (NIR). These studies typically solved the restricted NLTE problem for trace elements within 1D~LTE models (e.g.\,\citealt{Mashonkina2000, Takeda2003, Shi2004, Andrievsky2007, Lind2011}). The most comprehensive 1D~NLTE grid for \ion{Na}{I} is provided by \citet{Lind2022}, which includes NLTE effects for 35 \ion{Na}{I} lines spanning the ultra-violet (UV) to NIR. 
The most recent 1D~NLTE grid in the optical for metal-poor stars (i.e. [Fe/H]$\leq -2.0$) is published in \citet{Mashonkina2023}, with abundance corrections in very good agreement with those from \citet{Lind2022}, as shown in Fig.~9 of the former paper. 

The primary NLTE effect for \ion{Na}{I} lines in 1D model atmospheres is the deepening of the cores of strong lines, primarily due to photon suction and over-recombination. Over-population of the lower states of the transitions and sub-thermal source functions lead to stronger spectral lines in NLTE compared to LTE for the same sodium abundance. As a result, NLTE corrections based on equivalent widths are generally negative and can exceed $-0.5$\,dex compared to 1D LTE in cases of full line saturation (e.g.\,\citealt{Marino2011}). This highlights the necessity of applying NLTE corrections to derive accurate sodium abundances.

In order to validate our models, in Fig.~\ref{fig: comparison1D} we compare our 1D~NLTE and 1D~LTE reduced equivalent widths\footnote{The ratio of REWs in NLTE and LTE is approximately the same as the abundance correction (but with opposite sign) for weak lines.} with previous studies that employed, with minor modifications, the same model atom and \texttt{MARCS} atmospheres as adopted here. Specifically, our results are compared with those of \citet{Lind2011} (L11) and \citet{Lind2022} (L22) for four \ion{Na}{I} lines in a turn-off star (upper panels) and a giant (lower panels). 
In both L11 and L22, values below a prescribed minimum theoretical equivalent width were not computed and therefore do not appear in the figure. 

Overall, our models closely follow the L22 trends, with only small deviations at low [Fe/H]. The improved agreement with L22 relative to L11 is plausibly linked to methodological similarities. In L22, the spectrum synthesis was carried out with PySME \citep{Wehrhahn2023}\footnote{\url{sme.astro.uu.se}}, while the departure coefficients, i.e. the ratios of NLTE to LTE level populations as a function of atmospheric depth, were taken from \texttt{Balder}. This setup results in a treatment of background opacities and the equation of state similar to that used in this work, albeit with a slightly different model atom (lacking the modified H-collision rates introduced by \citealt{Canocchi2024a}; see Sect.~\ref{sec: modelatom}).
The residual discrepancies between our results and L22 may reflect these differences in the adopted hydrogen-collision prescriptions, particularly as the deviations appear to increase towards lower [Fe/H], although this should be regarded only as a possible explanation.

In contrast, L11 computed both the statistical-equilibrium solution (i.e. the departure coefficients) and the spectrum synthesis using the \texttt{MULTI2.3} code (\citealt{Carlsson1992}). The comparison therefore primarily highlights differences arising from the choice of code used to solve the statistical equilibrium and to perform spectrum synthesis. 
Indeed, L22 concluded that the discrepancies relative to L11 likely do not originate from differences in the statistical-equilibrium solutions themselves, but from the calculation of LTE and NLTE line and continuous opacities in \texttt{MULTI2.3} compared with PySME.

\subsection{Departures from LTE in 3D}\label{sec: 3Ddepartures}
In 3D, the NLTE effects vary significantly over the granulation pattern, with clear differences between the hot granules and the cool intergranular lanes, as illustrated in Fig.~\ref{fig: resolvedNa}. The figure shows the spatially resolved ratio of 3D~NLTE to 3D~LTE equivalent widths for six \ion{Na}{I} lines, computed at the surface of a representative \texttt{Stagger} snapshot of a metal-poor model atmosphere with $T_\mathrm{eff} = 4978.17$\,K, $\log g = 2.0$, [Fe/H]$ = -2.0$, and $A(\mathrm{Na}) = 4.22$. The spectrum synthesis was performed on a model atmosphere with resolution $120 \times 120 \times 120$ to provide a clearer visualisation of the NLTE–LTE differences across the convective surface.

For all lines except the \ion{Na}{I}~D doublet, the line strength is enhanced in NLTE (i.e. appears blue in the figure) across nearly the entire surface, reflecting the dominance of over-deexcitation and overrecombination both in the granules and the lanes. In contrast, the resonance lines at 5889\,\r{A} and 5896\,\r{A} behave differently. In the hotter granules, overionisation from the ground state becomes particularly efficient, leading to weaker NLTE line profiles (i.e. red). Conversely, in the cooler intergranular lanes, the same line strengthening is seen as for other lines.

Because the granules dominate the surface area, the net NLTE effect in 3D is a weakening of the \ion{Na}{I}~D lines compared to 3D~LTE. This behaviour can be understood as follows. The \ion{Na}{I}~D lines correspond to transitions from the ground state (3s) to the two fine-structure components of the first excited state (3p$_{1/2}$ and 3p$_{3/2}$). These excited levels have photoionisation thresholds around 400\,nm, so they are strongly affected by the superthermal UV radiation field ($J_\nu > B_\nu$), which lead to overionisation of \ion{Na}{I}. 
By contrast, more highly excited states that produce transitions of subordinate \ion{Na}{I} lines have photoionisation thresholds in the NIR, where instead $J_\nu < B_\nu$. 

The steep temperature gradient in the granules produces a significant $(J_\nu - B_\nu)$ excess, in particular for metal-poor stars. The overionisation from the first excited state underpopulates that state as well as the ground state. 
Indeed, under these conditions, we find that the lower and upper levels are close to relative LTE, meaning that their departure coefficients ($b\equiv n^\mathrm{NLTE}/n^\mathrm{LTE}$) satisfy $b_\mathrm{up}/b_\mathrm{low} \approx 1$, even though both levels are underpopulated with respect to LTE.
The resulting opacity shortage leads to weaker NLTE resonance lines at fixed abundance, exactly as seen in Fig.~\ref{fig: resolvedNa}. It is important to note that we only see this effect in 3D, because of the steeper temperature gradients. At the typical line-formation depth, the temperature contrast between granules and intergranular lanes can exceed 1500\,K in this model. The average formation depth is determined from the contribution function to the absolute flux depression, computed following Eq.~15 of \citet{Amarsi2015}. 

At low metallicity ([Fe/H]~$\leq -2.0$), this weakening becomes the dominant NLTE effect. The resulting abundance corrections between 3D~NLTE and 3D~LTE are therefore positive and can reach $+0.5$\,dex at line saturation, as shown in Fig.~\ref{fig: 3N3Labcorr5889}, where the abundance corrections in 3D as a function of the REW for all the models in our grid are shown.
At higher [Fe/H], the effects of UV overionisation are diminished. The \ion{Na}{I}~D lines are stronger and photon losses from the line core become apparent. As in the 1D case, the line source function is subthermal (i.e. weaker than in LTE). Combined with the overpopulation of the ground state, this leads to a substantial NLTE line strengthening and negative abundance corrections, consistent with the behaviour discussed for the other lines in the following section.

\subsection{3D NLTE abundance corrections}\label{sec: 3DNLTEcorr}
In order to quantify the NLTE effects in 3D, we compute abundance corrections, as defined in Eq.~\ref{eq:corrAb}, for each \ion{Na}{I} line, by interpolating the abundances in the 1D grids onto the stellar parameters ($T_\mathrm{eff}$, $\log g$, [Fe/H]), and REW of the 3D grid.

Figure~\ref{fig: ALLabcorr5688} displays the abundance corrections for each computed model and abundance compared to the REW of the other spectral models (1D~LTE, 1D~NLTE, and 3D~LTE) for the line at 5688\,\r{A}. In this way, it is possible to investigate how the abundance correction varies with line strength and saturation.
All the other \ion{Na}{I} lines show a very similar behaviour in shape to the 5688\,\r{A} line, except for the resonance lines (D lines) in 3D~NLTE vs 3D~LTE.

From this figure, it is clear that the largest abundance corrections occur when the line is close to saturation, which happens at a REW of about $-4.8$, as already noted by works on other elements in 3D~NLTE such as \ion{Ca}{II} (\citealt{Lagae2025}), \ion{Fe}{I} (\citealt{Amarsi2022}), and \ion{Ca}{I} and \ion{O}{I} (\citealt{Amarsi2019}). Indeed, since the \ion{Na}{I} lines saturate at different abundances depending on the assumption, 1D or 3D, LTE or NLTE, the largest difference between curve of growths occurs close to saturation. 

The difference between the 3D~NLTE and 1D~LTE models with a $v_\mathrm{mic}=1.0$\,km\,s$^{-1}$ is shown in the upper left panel. The 5688\,\r{A} line saturates first in 1D~LTE than in 3D~NLTE, leading to negative abundance corrections down to $\Delta^\mathrm{3N}_\mathrm{1L} \approx -0.40$\,dex. For strong lines, the corrections become positive, up to $+0.14$.
On the other hand, the corrections for weak lines are in the range $-0.13, +0.02$.
Considering the stellar surface gravity, the largest abundance corrections occur for giants with $\log{g}<2.5$.
Tests with other values of $v_\mathrm{mic}$ show a very similar pattern. An example for the 5682\,\r{A} line is shown in Fig.~\ref{fig: abcorrdiffvmic} in Appendix.

Most of the abundance corrections $\Delta^\mathrm{3N}_\mathrm{1L}$ for the other \ion{Na}{I} lines follow a similar pattern. The largest correction occurs for the 8194\,\r{A} and the \ion{Na}{I} D lines at saturation, with the corrections dropping to about $-0.69$ and $-0.68$, respectively (see Fig.~\ref{fig: ALLabcorr5889} and \ref{fig: ALLabcorr8194}). The only exception is the line at 10\,747\,\r{A} in the infrared, which remains consistently weak for all models in the [Na/Fe] range of $-0.5$ to 1.0, showing predominantly positive abundance corrections within the range of $-0.05$ to 0.10. The abundance corrections for this line are shown in Appendix in Fig.~\ref{fig: ALLabcorr10747}. 

In the other panels of Fig.~\ref{fig: ALLabcorr5688}, we also illustrate the 3D~NLTE corrections in comparison to the 1D~NLTE (top right panel) and 3D~LTE (bottom right panel) models. The former, $\Delta^\mathrm{3N}_\mathrm{1N}$, behaves differently from the 1D~LTE case. In fact, the 3D~NLTE lines saturate first, resulting in positive abundance corrections for most of the lines, again with a negative peak of $-0.1$\,dex near saturation, which is less severe than in the $\Delta^\mathrm{3N}_\mathrm{1L}$ case. Specifically, for weak lines, the corrections are mostly positive, in the range $\Delta^\mathrm{3N}_\mathrm{1N}= -0.01$ to $+0.09$.
Overall, the abundance corrections relative to 1D~LTE, $\Delta^\mathrm{3N}_\mathrm{1L}$, for weak lines are closer to zero than the corrections relative to 1D~NLTE, $\Delta^\mathrm{3N}_\mathrm{1N}$. This is caused by a cancellation effect between negative NLTE corrections and positive 3D corrections.

The abundance corrections between 3D~NLTE and 3D~LTE, $\Delta^\mathrm{3N}_\mathrm{3L}$, and 1D~NLTE and 1D~LTE, $\Delta^\mathrm{1N}_\mathrm{1L}$ in the bottom right and left panel of Fig.~\ref{fig: ALLabcorr5688}, respectively, show a very similar pattern. In both cases, the corrections are negative across the full range of line strengths, typically spanning $-0.03$ to $-0.31$\,dex.
In particular, $\Delta^\mathrm{3N}_\mathrm{3L}$ displays a similar trend for all \ion{Na}{I} lines, except for the D lines. For these resonance lines, overionisation in the granules of metal-poor stars becomes highly efficient, leading to positive abundance corrections, as discussed in the previous section.
It is important to note that this is the reverse of the 1D case.
This is particularly relevant because D lines are often the only \ion{Na}{I} lines detectable in metal-poor stars and thus are commonly used in sodium abundance studies in this regime.

In Fig.~\ref{fig: abcorr5688}, we show the variation of the abundance correction $\Delta^\mathrm{3N}_\mathrm{1L}$ across the Kiel diagram for the \ion{Na}{I} 5688\,\r{A} line, at a fixed [Na/Fe] ratio and four representative stellar metallicities.
At very low metallicities ([Fe/H]$=-4.0$), the 5688\,\r{A} line is extremely weak and nearly undetectable, resulting in negligible corrections close to zero.
For moderately metal-poor stars ([Fe/H]$=-2.0$ to $-1.0$), the corrections become increasingly negative with decreasing surface gravity at fixed $T_\mathrm{eff}$, reflecting the strengthening of NLTE effects in lower-density atmospheres. 
At solar metallicity ([Fe/H]$=0.0$), the line gradually transitions from the saturated to the strong regime as $T_\mathrm{eff}$ decreases, producing slightly positive abundance corrections at low effective temperatures ($T_\mathrm{eff} \approx 4000$–$4500$\,K).
Overall, the largest 3D~NLTE–1D~LTE corrections are found in stars with low $\log g$ and high $T_\mathrm{eff}$, where departures from LTE are most pronounced.
All the other subordinate lines exhibit a nearly identical trend across the parameter space, except for the $10\,747$\,\r{A} line, which consistently has slightly positive $\Delta^\mathrm{3N}_\mathrm{1L}$ at every $T_\mathrm{eff}$, $\log g$, and [Fe/H].
The behaviour of the \ion{Na}{I} D lines in the parameter space is shown in Fig.~\ref{fig: abcorr5889HR} in Appendix.

\subsection{Recommended lines for abundance determination in different stars}\label{sec: recommendations}
\ion{Na}{I} has few observable lines in late-type stellar spectra, and the available lines are often saturated.  
For instance, while the strong \ion{Na}{I} D lines are not ideal for precise abundance determination via $W_\lambda$ analysis, they may be the only lines detectable in warm, metal-poor dwarf stars. In such cases, a detailed line profile fitting, especially of the line wings, is more suitable than an equivalent width analysis for obtaining reliable abundance estimates.

When measuring equivalent widths of \ion{Na}{I} lines, it is generally preferable to use unsaturated lines.  
The equivalent widths of such lines are more sensitive to small changes in abundance, thus facilitating more precise abundance determinations. They are also less sensitive to departures from LTE, as NLTE corrections tend to be smaller (see Sect.~\ref{sec: 3DNLTEcorr}).
The impact of 3D~NLTE abundance corrections on different \ion{Na}{I} lines for the same star is illustrated in Fig.~\ref{fig: 2metpoor}, which shows examples of a solar-metallicity dwarf and a metal-poor giant. In both cases, the corrections are minimal for the 6154\,\r{A} line (typically between $-0.1$ and 0.0\,dex), although this line may become too weak to detect in metal-poor stars. 

The 5682\,\r{A} line is somewhat stronger (i.e. has a larger REW) and exhibits similarly small abundance corrections in the metal-poor case. However, the corrections become increasingly negative toward higher metallicities, as the line approaches saturation. In contrast, the lines at 8194\,\r{A} and 5889\,\r{A} are very strong in both stars. The 8194\,\r{A} line shows large negative corrections in both cases, reaching down to $-0.75$\,dex in the metal-poor giant. The 5889\,\r{A} (Na~D$_2$) line, though saturated, displays a strong metallicity dependence: in the metal-poor giant, its abundance correction can be as large as $-0.6$\,dex, whereas in the solar-metallicity dwarf it is relatively small (ranging from $-0.26$ to $-0.03$\,dex), as the line transitions from the saturated regime toward the damping part of the curve of growth.

In general, in stars with [Fe/H]$\leq -2.5$, the 5682/5688\,\r{A} doublet lines are typically too weak to be observed, but they serve as good abundance indicators for stars with [Fe/H] between $-2.0$ and $-1.0$. At higher metallicities, including solar and super-solar values, the 6154/6160\,\r{A} lines are to be preferred.

Finally, in very metal-poor stars, we recommend that $W_\lambda$ of the resonance \ion{Na}{I} D lines should be used only as a last resort—specifically when the 8183/8194\,\r{A} doublet is either undetectable or heavily contaminated by telluric absorption. This is owing to their extremely severe abundance corrections as discussed above.

\subsection{Validation on benchmark stars}\label{sec: benchmark}
We validated our 3D~NLTE grid of synthetic \ion{Na}{I} spectra using high-resolution optical observations of five well-established benchmark stars: the Sun, Arcturus, HD~84937, HD~140283, and HD~122563. 
For most lines, equivalent widths were adopted from L22. For the 5682/5688\,\r{A} doublet in HD~84937 and HD~140283, however, we used high-resolution ESPRESSO/VLT spectra (\citealt{Pepe2021}) with S/N~$> 1000$ from \citet{Wang2022}, measuring the equivalent widths following the same procedure as in L22. Further details of this dataset are provided in Sect.~2 of \citet{Wang2022}.
The stellar parameters listed in Table~\ref{tab: benchmarks} were used to derive $A(\mathrm{Na})$ under different modelling assumptions (1D/3D; LTE/NLTE). Additional details on the adopted stellar parameters are provided in Sect.~3.3 of L22.

Figure~\ref{fig: benchmarks} presents the resulting $A(\mathrm{Na})$ values for each star, computed as the weighted mean of the available lines with $W_\lambda/\lambda < 3 \times 10^{-5}$ (corresponding to REW$\approx -4.52$), following L22. We note, however, that a more restrictive REW threshold—for example, REW $\leq -5$, which selects only weak lines on the linear part of the curve of growth—may be preferable in some applications, depending on the specific abundance analysis goals.  
In addition to the five benchmark stars, we included the ultra metal-poor star SDSS J102915.14+172927.9. For this star, the \ion{Na}{I} abundance is derived as an upper limit from the equivalent width of the \ion{Na}{I} D$_2$ line, following \citet{Lagae2023}. This upper limit lies slightly below the lower boundary of our grid and was therefore obtained by extrapolation. Nevertheless, it is in very good agreement with the values reported by \citet{Lagae2023}, with a difference of less than 0.01\,dex for the 3D~NLTE abundance, and less than 0.04\,dex for the other models.

For the Sun and Arcturus, all nine \ion{Na}{I} lines included in our grid were available. In contrast, for HD~84937 and HD~140283, only the strong \ion{Na}{I} D lines, the 5682/5688\,\r{A} doublet, and the 8194\,\r{A} line were measured. For HD~122563, the dataset comprises the \ion{Na}{I} D lines as well as the 5682/5688\,\r{A} doublet. We note that HD~122563, with $\log g =1.40$, lies slightly below the lower boundary of our grid ($\log g=1.50$), and so the abundance corrections for this star are based on extrapolation. The line-by-line measured equivalent widths and derived abundances are reported in Table~\ref{tab: linebylineANa} in Appendix.

In all cases, the line-to-line scatter is reduced in 3D~NLTE compared to both LTE models (1D and 3D). While the reduction relative to 1D~LTE is modest for the Sun and Arcturus ($\sim 0.03$\,dex), it becomes significant for the metal-poor stars ($\geq 0.1$\,dex). 
A smaller yet noticeable improvement is also found in 3D~NLTE with respect to 1D~NLTE, with the scatter further reduced by up to $\leq 0.02$\,dex.

For the benchmark stars, the 1D models yield mean \ion{Na}{I} abundances consistent with those reported in L22, with all differences remaining within the $1\sigma$ standard deviation. The average offset amounts to about 0.06\,dex in 1D~LTE and 0.03\,dex in 1D~NLTE.
These small discrepancies can be attributed primarily to differences in the set of lines used for the abundance determination—L22 includes a significantly larger number of transitions—as well as to the adopted fudge parameter for the microturbulence ($v_\mathrm{mic}$). While L22 employed the $v_\mathrm{mic}$ values listed in their Table~2, we adopt a fixed value of $v_\mathrm{mic}=1.0$\,km\,s$^{-1}$ in this work, which may also contribute to the residual differences.

The 3D~NLTE abundance corrections are generally negative relative to 1D~LTE. An exception is the ultra metal-poor star, for which the correction is positive for the \ion{Na}{I} D$_2$ line (as also found by \citealt{Lagae2023}), leading to a higher inferred abundance and thus behaving in the opposite direction compared to the 1D~NLTE correction.

\section{Conclusions}\label{sec: conclusions}
In this work, we investigate the impact of \ion{Na}{I} line formation in 3D~NLTE on the determination of stellar sodium abundances in late-type FGK stars. For this purpose, we compute a grid of synthetic stellar spectra for nine commonly used \ion{Na}{I} lines using 3D RHD stellar models from the recently published \texttt{Stagger} grid, and NLTE line formation. 
The 3D~NLTE grid builds upon the one presented in \citet{Canocchi2024b} for dwarfs, and now extends its coverage to include giants and metal-poor stars, spanning the following range of stellar parameters:
\begin{itemize}
    \item Effective temperature: between 4000 and 6500\,K, in steps of about 500\,K. Note that the $T_\mathrm{eff}$ points are slightly irregular;
    \item Surface gravity: between 1.5 and 5.0, in steps of 0.5\,dex;
    \item Metallicity: $-4.0, -3.0, -2.0, -1.0, 0.0, +0.5$\,dex;
    \item Sodium abundance ([Na/Fe]): from $-0.5$ to $+0.5$ for the dwarfs, and $-0.5$ to $+1.0$ for the giants ($\log g \leq 3.5$), in steps of 0.5\,dex.
\end{itemize}
We trained radial basis functions to interpolate line profiles through the grid at any given set of stellar parameters ($T_\mathrm{eff}$, $\log g$, and [Fe/H]) and sodium abundance. Line profiles in 1D~LTE, 1D~NLTE, and 3D~LTE are computed as well for comparison with the 3D~NLTE models. These grids are then used to calculate abundance corrections for every sodium line for a range of equivalent width values. We trained fully connected feed-forward neural networks to predict sodium abundances over the grid from a given set of stellar parameters ($T_\mathrm{eff}$, $\log g$, and [Fe/H]) and the reduced equivalent widths. Leave-one-out cross-validation shows that the estimates of the interpolation errors are small, with a median absolute deviation below $\sim$0.01\,dex for most \ion{Na}{I} lines, and only slightly higher for the D lines ($\sim$0.015\,dex).

Validation against a set of benchmark stars indicates that the line-to-line scatter decreases in 3D~NLTE compared to the 1D~LTE case. The corresponding abundance corrections are generally negative, though less pronounced than in 1D~NLTE. An exception is found for the \ion{Na}{I} D lines in ultra metal-poor stars, which exhibit positive corrections relative to 1D~LTE.

We conclude that the 3D~NLTE abundance corrections with respect to 1D~LTE are mostly negative. For unsaturated lines, 1D~NLTE performs quite well, with 3D~NLTE abundances up to about 0.1\,dex higher. However, for saturated and strong lines, abundance correction can be significant, with variations down to $-0.7$\,dex for the \ion{Na}{I} D lines, which are often the only available diagnostic in very metal-poor stars.

We make our 3D~NLTE grid, along with the associated interpolation routines, publicly available to facilitate more accurate sodium abundance determinations in current and future stellar spectroscopic surveys. In a forthcoming paper (Canocchi et al., in prep.), we apply this grid to the almost one million Milky Way stars from GALAH DR4 (\citealt{GALAHDR4}), deriving for the first time homogeneous 3D~NLTE \ion{Na}{I} abundances across a broad range of stellar parameters and Galactic environments. This work will represent the first 3D~NLTE analysis of Na applied to a large spectroscopic survey, providing improved constraints for Galactic archaeology and chemical evolution studies.



\begin{acknowledgements} 
We thank the anonymous referee for their comments, which have improved the manuscript.
GC and KL acknowledge funds from the Knut and Alice Wallenberg foundation. KL and EXW also acknowledge funds from the European Research Council (ERC) under the European Union’s Horizon 2020 research and innovation programme (Grant agreement No. 852977).
AMA acknowledges support from the Swedish Research Council (VR 2020-03940, VR 2025-05167) the Crafoord Foundation via the Royal Swedish Academy of Sciences (CR 2024-0015), and the European Union’s Horizon Europe research and innovation programme under grant agreement No. 101079231 (EXOHOST).
We thank the PDC Center for High Performance Computing, KTH Royal Institute of Technology, Sweden, for providing access to computational resources and support. 
The computations were enabled by resources provided by the National Academic Infrastructure for Supercomputing in Sweden (NAISS), partially funded by the Swedish Research Council through grant agreement no. 2022-06725, at the PDC Center for High Performance Computing, KTH Royal Institute of Technology (project numbers NAISS 2023/1-15 and NAISS 2024/1-14).
This research has made use of NASA's Astrophysics Data System (ADS) bibliographic services. 
We acknowledge the community efforts devoted to the development of the following open-source packages that were used in this work: numpy (\url{numpy.org}), matplotlib (\url{matplotlib.org}), and astropy (\url{astropy.org}). 

\end{acknowledgements}

\section*{Data Availability}
The grid of 3D~NLTE synthetic spectra and abundance corrections for \ion{Na}{I} described in Sect.~\ref{sec: synthspectra}, together with the interpolation routines described in Sect.~\ref{sec: interpolation} can be downloaded at \url{https://doi.org/10.5281/zenodo.19201830}.

\bibliographystyle{aa_url}
\bibliography{references} 

\onecolumn
\appendix 
\section{Additional tables and figures} 
\begin{table*}[h]
\centering 
\caption{Parameters of the 3D~NLTE synthetic spectra computed in the grid for this work.}\label{tab: synthspectraparams} 
\begin{tabular}{l c c c} \hline \hline
\noalign{\smallskip}
$\mathrm{[Fe/H]}$ & $\log{g}$ & $T_\mathrm{eff}$ & $A$(Na)$^{(a)}$ \\ 
{[dex]} &  {[cm s$^{-2}$]} & [K] & [dex] \\ \hline 
 \noalign{\smallskip}
$-4.0$ & 1.5 & 4000, 4500 & 1.72, 2.22, 2.72, 3.22 \\
 & 2.0 & 4500, 5000 & " \\
 & 2.5 & 4000, 4500, 5000 & " \\
 & 3.0 & 4500, 5000, 5500 & " \\
 & 3.5 & 4500, 5000, 5500, 6000 & " \\
 & 4.0 & 5000, 5500, 6000, 6500 & 1.72, 2.22, 2.72 \\
 & 4.5 & 4500, 5000, 5500, 6000, 6500 & " \\
 & 5.0 & 5000, 5500 & " \\
$-3.0$ & 1.5 & 4000, 4500 & 2.72, 3.22, 3.72, 4.22 \\ 
 & 2.0 & 4000, 4500, 5000 & " \\
 & 2.5 & 4000, 4500, 5000, 5500 & " \\
 & 3.0 & 4500, 5000, 5500 & " \\
 & 3.5 & 4500, 5000, 5500, 6000 & " \\
 & 4.0 & 4500, 5000, 5500, 6000, 6500 & 2.72, 3.22, 3.72 \\
 & 4.5 & 5000, 5500, 6000, 6500 & " \\
 & 5.0 & 4500, 5000 & " \\
$-2.0$ & 1.5 & 4000, 4500 & 3.72, 4.22, 4.72, 5.22 \\
 & 2.0 & 4000, 4500, 5000, & " \\
 & 2.5 & 4000, 4500, 5000, 5500 & " \\
 & 3.0 & 4500, 5000, 5500 & " \\
 & 3.5 & 4500, 5000, 5500, 6000 & " \\
 & 4.0 & 4500, 5000, 5500, 6000, 6500 & 3.72, 4.22, 4.72 \\
 & 4.5 & " & " \\
 & 5.0 & 4500, 5000, 5500 & " \\
$-1.0$ & 1.5 & 4000, 4500 & 4.72, 5.22, 5.72, 6.22 \\
 & 2.0 & 4000, 4500, 5000 & " \\
 & 2.5 & 4000, 4500, 5000, 5500 & " \\
 & 3.0 & 4500, 5000, 5500 & " \\
 & 3.5 & 4500, 5000, 5500, 6000 & " \\
 & 4.0 & 4500, 5000, 5500, 6000, 6500 & 4.72, 5.22, 5.72 \\
 & 4.5 & 5000, 6000, 6500 & " \\
 & 5.0 & 4500, 5000, 5500 & " \\
$0.0$ & 1.5 & 4000, 4500 & 5.72, 6.22, 6.72, 7.22 \\
 & 2.0 & 4000, 4500, 5000 & " \\
 & 2.5 & 4000, 4500, 5000 & " \\
 & 3.0 & 4500, 5000, 5500 & " \\
 & 3.5 & 4500, 5000, 5500, 6000 & " \\
 & 4.0 & 4500, 5000, 5500, 6000, 6500 & 5.72, 6.22, 6.72 \\
 & 4.5 & 4000, 4500, 5000, 5500, 6000, 6500 & " \\
 & 5.0 & 4500, 5000, 5500 & " \\
$+0.5$ & 1.5 & 4000 & 6.22, 6.72, 7.22, 7.72 \\
 & 2.0 & " & " \\
 & 2.5 & 4000, 4500 & " \\
 & 3.0 & 4500, 5000, 5500 & " \\
 & 3.5 & 4500, 5000, 5500 & " \\
 & 4.0 & 4500, 5000, 5500, 6000, 6500 & 6.22, 6.72, 7.22 \\
 & 4.5 & 4500, 5000, 5500, 6000, 6500 & " \\
 & 5.0 & 4500, 5000, 5500 & " \\ 
\hline 
\end{tabular}\\
\begin{flushleft}
\textbf{Notes:}\\
$^{(a)}$ We take as the reference for the solar abundance the value reported in \citet{Asplund2021}: $A\mathrm{(Na)}_\odot = 6.22 \pm 0.03$. In the relative abundance notation, this corresponds to [Na/Fe]=0.0 for the solar metallicity models (i.e. [Fe/H]=0.0).

\end{flushleft}
\end{table*}
\begin{table} 
    \centering
    \caption{Error statistics of the FFNN in predicting abundance ($A\text{(Na)}$) in the leave-one-out cross-validation on the grid in 1D~LTE, 1D~NLTE, and 3D~LTE. The columns show the root-mean-square (RMS) and the median absolute deviation (MAD) of the corresponding FFNN with optimized hyperparameters (number of layers, $n_l$, neurons per layer, $n$, and L2 penalty, $\alpha$) for different \ion{Na}{I} lines.}
    \label{tab: FFNNothermodels}
    \begin{tabular}{l c c c c c c} \hline \hline
    \noalign{\smallskip}
    $\lambda$ & model & RMS$({\Delta A\mathrm{(Na)}})$ & MAD$({\Delta A\mathrm{(Na)})}$ & $n_l$ & $n$ & $\alpha$ \\ 
     {[}\AA{]} & & [dex] & [dex] & & & \\ \hline 
    \noalign{\smallskip} 
     5682  & 1D LTE & 0.029 & 0.015 & 2 & 100 & 0.01\\ 
      & 1D NLTE & 0.019 & 0.009 & 2 & 100 & 0.001\\ 
      & 3D LTE & 0.028 & 0.014 & 2 & 300 & 0.1 \\
     5688 & 1D LTE & 0.026 & 0.011 & 2 & 500 & 0.001\\ 
      & 1D NLTE & 0.021 & 0.011 & 2 & 500 & 0.001\\ 
      & 3D LTE &  0.021 & 0.008 & 3 & 400 & 0.001 \\ 
     5889 & 1D LTE & 0.043 & 0.014 & 2 & 400 & 0.0001 \\ 
      & 1D NLTE & 0.040 & 0.019 & 2 & 400 & 0.0001 \\ 
      & 3D LTE & 0.036 & 0.016 & 2 & 400 & 0.01 \\ 
     5896 & 1D LTE & 0.039 & 0.014 & 2 & 100 & 0.0001 \\ 
      & 1D NLTE & 0.036 & 0.015 & 2 & 100 & 0.0001 \\ 
      & 3D LTE & 0.037 & 0.016 & 2 & 500 & 0.001 \\
     6154 & 1D LTE & 0.019 & 0.009 & 2 & 400 & 0.001 \\ 
      & 1D NLTE & 0.016 & 0.007 & 3 & 200 & 0.0001\\
      & 3D LTE & 0.026 & 0.011 & 3 & 100 & 0.0001\\ 
     6160 & 1D LTE & 0.034 & 0.017 & 2 & 500 & 0.01 \\ 
      & 1D NLTE & 0.019 & 0.009 & 2 & 200 & 0.001\\ 
      & 3D LTE & 0.026 & 0.009 & 2 & 300 & 0.0001 \\ 
     8183 & 1D LTE & 0.024 & 0.009 & 2 & 400 & 0.0001 \\ 
      & 1D NLTE & 0.036 & 0.018 & 2 & 500 & 0.01 \\ 
      & 3D LTE & 0.029 & 0.011 & 2 & 300 & 0.0001 \\
     8194 & 1D LTE & 0.041 & 0.019 & 3 & 200 & 0.01\\ 
      & 1D NLTE & 0.029 & 0.011 & 2 & 200 & 0.001\\ 
      & 3D LTE & 0.034 & 0.015 & 3 & 100 & 0.0001\\ 
     10\,747 & 1D LTE & 0.031 & 0.009 & 3 & 400 & 0.001 \\ 
     & 1D NLTE & 0.030 & 0.008 & 2 & 400 & 0.0001 \\ 
      & 3D LTE & 0.022 & 0.006 & 2 & 200 & 0.0001 \\ \hline 
      
    \end{tabular}
\end{table}
\begin{table}
    \centering
    \caption{Equivalent widths and derived sodium abundances ($A\mathrm{(Na)}$) with different models (1D/3D, LTE/NLTE) for all the lines available in the verification stars. In the 1D models, a $v_\mathrm{mic}=1.0$\,km\,s$^{-1}$ was adopted.}
    \label{tab: linebylineANa}
    \begin{tabular}{l c c c c c c} \hline \hline
    \noalign{\smallskip}
    & Sun & Arcturus & HD 84937 & HD 140283 & HD 122563 & SDSS J102915.14$+$172927.9\\ 
    $\lambda$ [\r{A}]& $W_\lambda$[m\r{A}] & $W_\lambda$[m\r{A}] & $W_\lambda$[m\r{A}] & $W_\lambda$[m\r{A}] & $W_\lambda$[m\r{A}] & $W_\lambda$[m\r{A}]\\
    & $A\mathrm{(Na)_{1DLTE}}$ & $A\mathrm{(Na)_{1DLTE}}$ & $A\mathrm{(Na)_{1DLTE}}$ & $A\mathrm{(Na)_{1DLTE}}$ & $A\mathrm{(Na)_{1DLTE}}$ & $A\mathrm{(Na)_{1DLTE}}$\\
    & $A\mathrm{(Na)_{1DNLTE}}$ & $A\mathrm{(Na)_{1DNLTE}}$ & $A\mathrm{(Na)_{1DNLTE}}$ & $A\mathrm{(Na)_{1DNLTE}}$ & $A\mathrm{(Na)_{1DNLTE}}$ & $A\mathrm{(Na)_{1DNLTE}}$ \\
    & $A\mathrm{(Na)_{3DLTE}}$ & $A\mathrm{(Na)_{3DLTE}}$ & $A\mathrm{(Na)_{3DLTE}}$ & $A\mathrm{(Na)_{3DLTE}}$ & $A\mathrm{(Na)_{3DLTE}}$ & $A\mathrm{(Na)_{3DLTE}}$\\
    & $A\mathrm{(Na)_{3DNLTE}}$ & $A\mathrm{(Na)_{3DNLTE}}$ & $A\mathrm{(Na)_{3DNLTE}}$ & $A\mathrm{(Na)_{3DNLTE}}$ & $A\mathrm{(Na)_{3DNLTE}}$ & $A\mathrm{(Na)_{3DNLTE}}$\\ \hline 
    \noalign{\smallskip} 
     5682 & $94.4 \pm 3.4$ & $131.3 \pm 5.4$ & $1.33 \pm 0.04$ & $0.87 \pm 0.04$ & $2.6 \pm 0.1$ & ... \\
     & $6.12 \pm 0.03$ & $6.12 \pm 0.08$ & $3.98 \pm 0.01$ & $3.58 \pm 0.02$ & $3.45 \pm 0.02$ & ...\\ 
     & $6.03 \pm 0.04$ & $5.91 \pm 0.08$ & $3.87 \pm 0.01$ & $3.47 \pm 0.01$ & $3.41 \pm 0.02$ & ...\\ 
     & $6.16 \pm 0.04$ & $6.21 \pm 0.07$ & $4.05 \pm 0.01$ & $3.63 \pm 0.02$ & $3.44 \pm 0.02$ & ...\\
     & $6.10 \pm 0.03$ & $6.02 \pm 0.08$ & $3.97 \pm 0.01$ & $3.57 \pm 0.02$ & $3.43 \pm 0.02$ & ...\\ 
     5688 & $134.3 \pm 2.8$ & $151.7 \pm 2.2$ & $2.45 \pm 0.05$ & $1.51 \pm 0.05$ & $4.1 \pm 0.2$ & ...\\
     & $6.17 \pm 0.02$ & $6.07 \pm 0.03$ & $3.99 \pm 0.01$ & $3.55 \pm 0.01$ & $3.40 \pm 0.02$ & ...\\
     & $6.07 \pm 0.02$ & $5.89 \pm 0.03$ & $3.90 \pm 0.01$ & $3.47 \pm 0.01$ & $3.34 \pm0.02$ & ...\\
     & $6.19 \pm 0.02$ & $6.20 \pm 0.03$ & $4.03 \pm 0.01$ & $3.60 \pm 0.01$ & $3.40 \pm 0.02$ & ...\\ 
     & $6.13 \pm 0.02$ & $5.99 \pm 0.03$ & $3.96 \pm 0.01$ & $3.53 \pm 0.01$ & $3.33 \pm 0.02$ & ...\\
     5889 & $766.4 \pm 8.0$ & $980.6\pm33.8$ & $113.5 \pm 0.7$ & $117.0\pm 3.0$ & $185.9 \pm 5.3$ & $5.5 \pm 1.0$ \\ 
     & $6.14 \pm 0.01$ & $5.55 \pm 0.03$ & $4.33 \pm 0.01$ & $3.95 \pm 0.05$ & $3.93 \pm 0.06$ & $1.56 \pm 0.10$\\
     & $6.05 \pm 0.01$ & $5.52 \pm 0.03$ & $3.76 \pm 0.01$ & $3.44 \pm 0.03$ & $3.58 \pm 0.07$ & $1.50 \pm 0.10$ \\
     & $6.18 \pm 0.01$ & $5.61 \pm 0.03$ & $3.72 \pm 0.01$ & $3.31 \pm 0.03$ & $3.35 \pm 0.06$ & $1.54 \pm 0.10$ \\
     & $6.14 \pm 0.01$ & $5.61 \pm 0.03$ & $3.88 \pm 0.01$ & $3.55 \pm 0.03$ & $3.57 \pm 0.06$ & $1.62 \pm 0.10$\\
     5896 & $557.9 \pm 6.3$ & $678.3 \pm 12.9$ & $94.0 \pm 0.6$ & $90.5 \pm 0.5$ & $161.0 \pm 2.4$ & ...\\
     & $6.11 \pm 0.01$ & $5.47 \pm 0.02$ & $4.30 \pm 0.01$ & $3.80 \pm 0.01$ & $3.91 \pm 0.03$ & ...\\
     & $6.01 \pm 0.01$ & $5.45 \pm 0.02$ & $3.85 \pm 0.01$ & $3.44 \pm 0.01$ & $3.52 \pm 0.03$ & ...\\
     & $6.14 \pm 0.01$ & $5.54 \pm 0.02$ & $3.78 \pm 0.01$ & $3.29 \pm 0.01$ & $3.28 \pm 0.03$ & ...\\
     & $6.08 \pm 0.01$ & $5.51 \pm 0.02$ & $3.94 \pm 0.01$ & $3.55 \pm 0.01$ & $3.49 \pm 0.03$& ...\\
     6154 & $38.8 \pm 0.5$ & $73.0 \pm 1.0$ & ... & ... & ... & ...\\
     & $6.26 \pm 0.01$ & $5.93 \pm 0.02$ & ... & ... & ... & ...\\
     & $6.18 \pm 0.01$ & $5.76 \pm 0.02$ & ... & ... & ... & ...\\
     & $6.28 \pm 0.01$ & $6.07 \pm 0.02$ & ... & ... & ... & ...\\
     & $6.21 \pm 0.01$ & $5.88 \pm 0.02$ & ... & ... & ... & ...\\
     6160 & $59.7 \pm 0.4$ & $98.2 \pm 0.7$ & ... & ... & ... & ...\\
     & $6.25 \pm 0.01$ & $6.08 \pm 0.01$ & ... & ... & ... & ...\\
     & $6.16 \pm 0.01$ & $5.86 \pm 0.01$ & ... & ... & ... & ...\\
     & $6.26 \pm 0.01$ & $6.18 \pm 0.01$ & ... & ... & ... & ...\\
     & $6.17 \pm 0.01$ & $5.98 \pm 0.01$ & ... & ... & ... & ...\\
     8183 & $229.6 \pm 1.8$ & $243.6 \pm 1.8$ & ... & ... & ... & ...\\
     & $6.29 \pm 0.01$ & $6.05 \pm 0.01$ & ... & ... & ... & ...\\ 
     & $6.00 \pm 0.01$ & $5.73 \pm 0.01$ & ... & ... & ... & ...\\
     & $6.32 \pm 0.01$ & $6.16 \pm 0.01$ & ... & ... & ... & ...\\
     & $6.07 \pm 0.01$ & $5.92 \pm 0.01$ & ... & ... & ... & ...\\
     8194 & $304.0 \pm 3.9$ & $298.3 \pm 2.4$ & $21.5 \pm 0.3$ & $13.1\pm0.4$ & ... & ...\\
     & $6.26 \pm 0.01$ & $6.02 \pm 0.02$ & $4.06 \pm 0.01$ & $3.56 \pm 0.02$ & ... & ...\\
     & $5.99 \pm 0.01$ & $5.78 \pm 0.02$ & $3.84 \pm 0.01$ & $3.39 \pm 0.02$ & ... & ...\\
     & $6.28 \pm 0.02$ & $6.12 \pm 0.01$ & $4.04 \pm 0.01$ & $3.52 \pm 0.02$ & ... & ... \\
     & $6.13 \pm 0.01$ & $5.95 \pm 0.01$ & $3.89 \pm 0.01$ & $3.50 \pm 0.02$ & ... & ... \\ 
     10\,747 & $13.2 \pm 1.0$ & $16.2 \pm 0.2$ & ... & ... & ... & ...\\
     & $6.21 \pm 0.04$ & $5.77 \pm 0.01$ & ... & ... & ... & ...\\
     & $6.20 \pm 0.04$ & $5.72 \pm 0.01$ & ... & ... & ... & ...\\
     & $6.20 \pm 0.04$ & $5.85 \pm 0.01$ & ... & ... & ... & ...\\ 
     & $6.21 \pm 0.04$ & $5.78 \pm 0.01$ & ... & ... & ... & ...\\
     \hline
     \end{tabular}
\end{table}
\begin{figure*}
    \centering
    \includegraphics[width=0.8\linewidth]{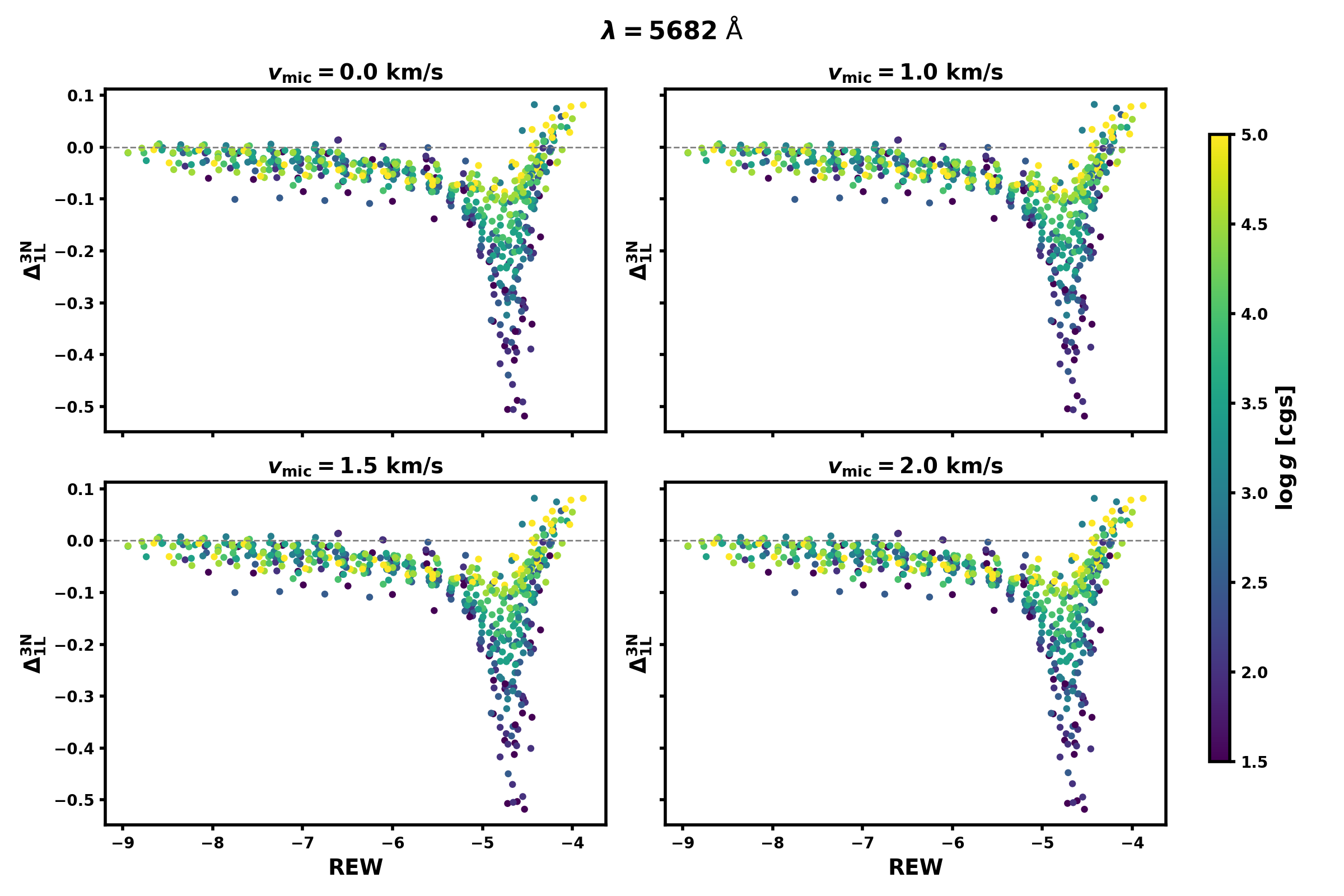}
    \caption{Abundance corrections (3D~NLTE to 1D~LTE) for the \ion{Na}{I} line at $5682$\,\r{A} for 1D models at $v_\mathrm{mic}=0.0,\,1.0,\,1.5,\,2.0$\,km\,s$^{-1}$, colour-coded as their $\log{g}$. The datapoints are computed at [Na/Fe]$=-0.5$ to $+0.5$ for the dwarfs, and $+1.0$ for the giants, in steps of 0.5\,dex} 
    \label{fig: abcorrdiffvmic}
\end{figure*}
\begin{figure*}
    \centering
    \includegraphics[width=0.8\linewidth]{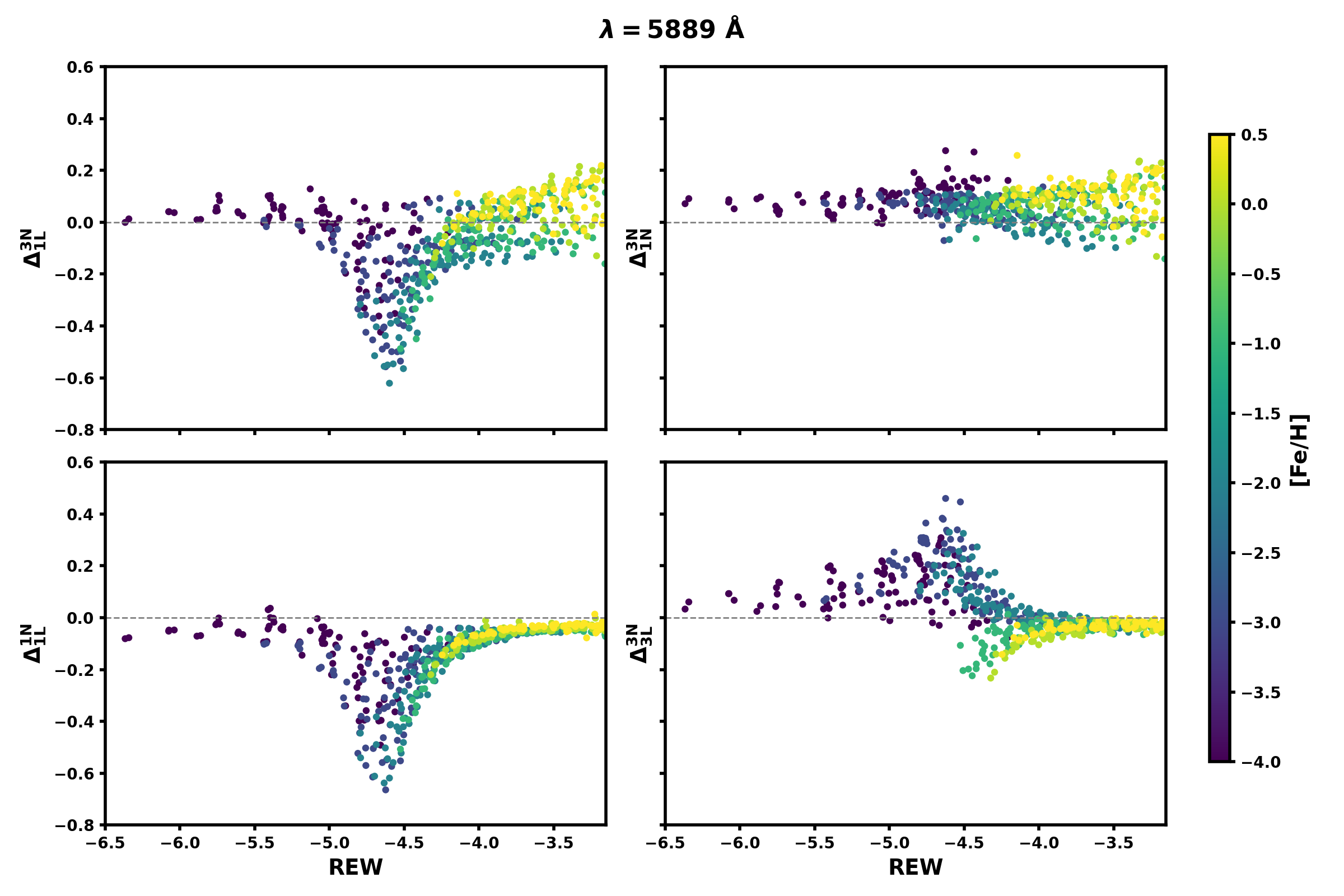}
    \caption{Abundance corrections for the \ion{Na}{I} line at $5889$\,\r{A} between different models: 3D~NLTE to 1D~LTE (top left), 3D~NLTE to 1D~NLTE (top right), 1D~NLTE to 1D~LTE (bottom left), and 3D~NLTE to 3D~LTE (bottom right), colour-coded as their [Fe/H]. The datapoints are computed at [Na/Fe]$=-0.5$ to $+0.5$ for the dwarfs, and $+1.0$ for the giants, in steps of 0.5\,dex and, for the 1D models, at a $v_\mathrm{mic}=1.0$\,km s$^{-1}$.} 
    \label{fig: ALLabcorr5889}
\end{figure*}
\begin{figure*}
    \centering
    \includegraphics[width=0.8\linewidth]{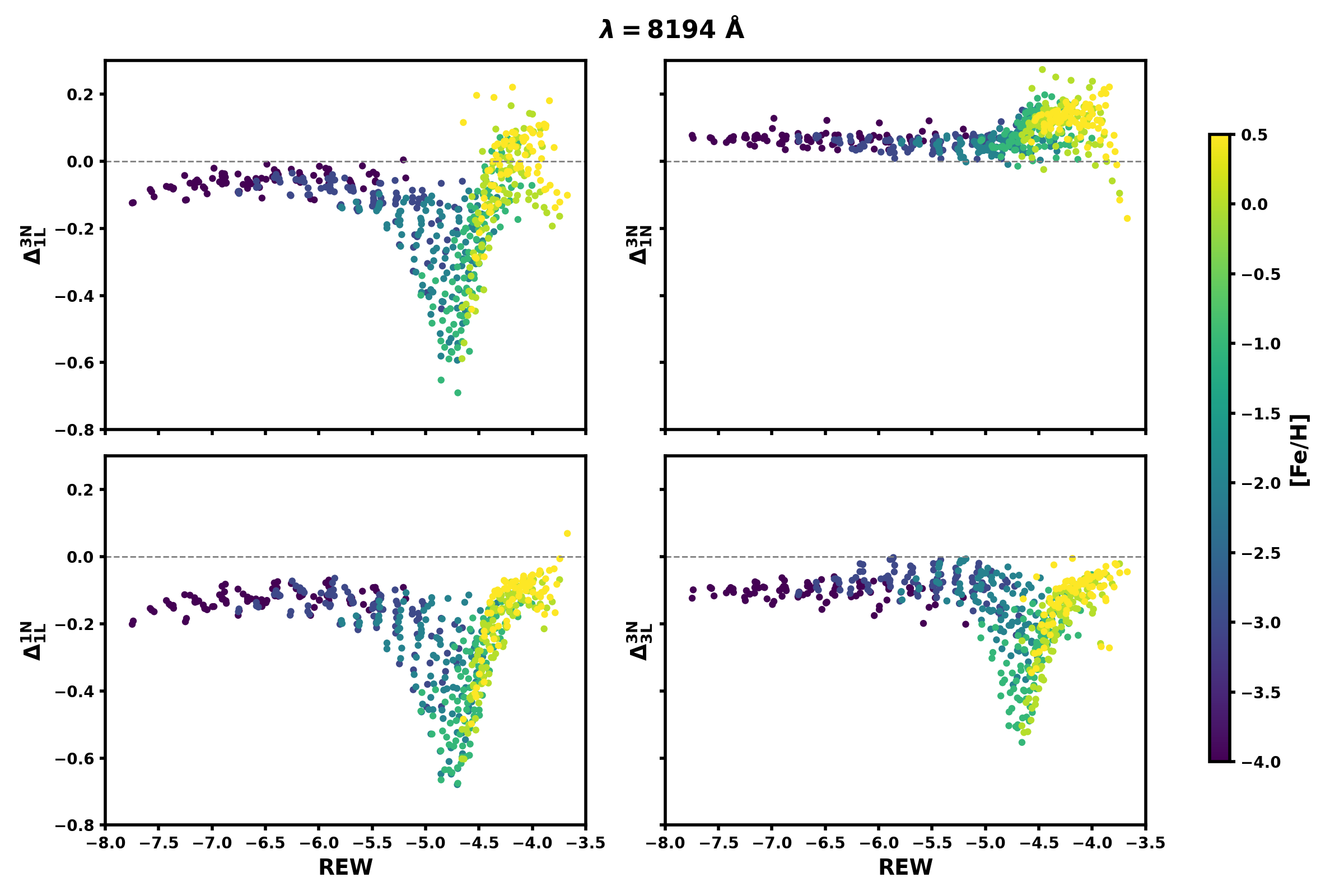}
    \caption{Abundance corrections for the \ion{Na}{I} line at $8194$\,\r{A} between different models: 3D~NLTE to 1D~LTE (top left), 3D~NLTE to 1D~NLTE (top right), 1D~NLTE to 1D~LTE (bottom left), and 3D~NLTE to 3D~LTE (bottom right), colour-coded as their [Fe/H]. The datapoints are computed at [Na/Fe]$=-0.5$ to $+0.5$ for the dwarfs, and $+1.0$ for the giants, in steps of 0.5\,dex and, for the 1D models, at a $v_\mathrm{mic}=1.0$\,km s$^{-1}$.} 
    \label{fig: ALLabcorr8194}
\end{figure*}
\begin{figure*}
    \centering
    \includegraphics[width=0.8\linewidth]{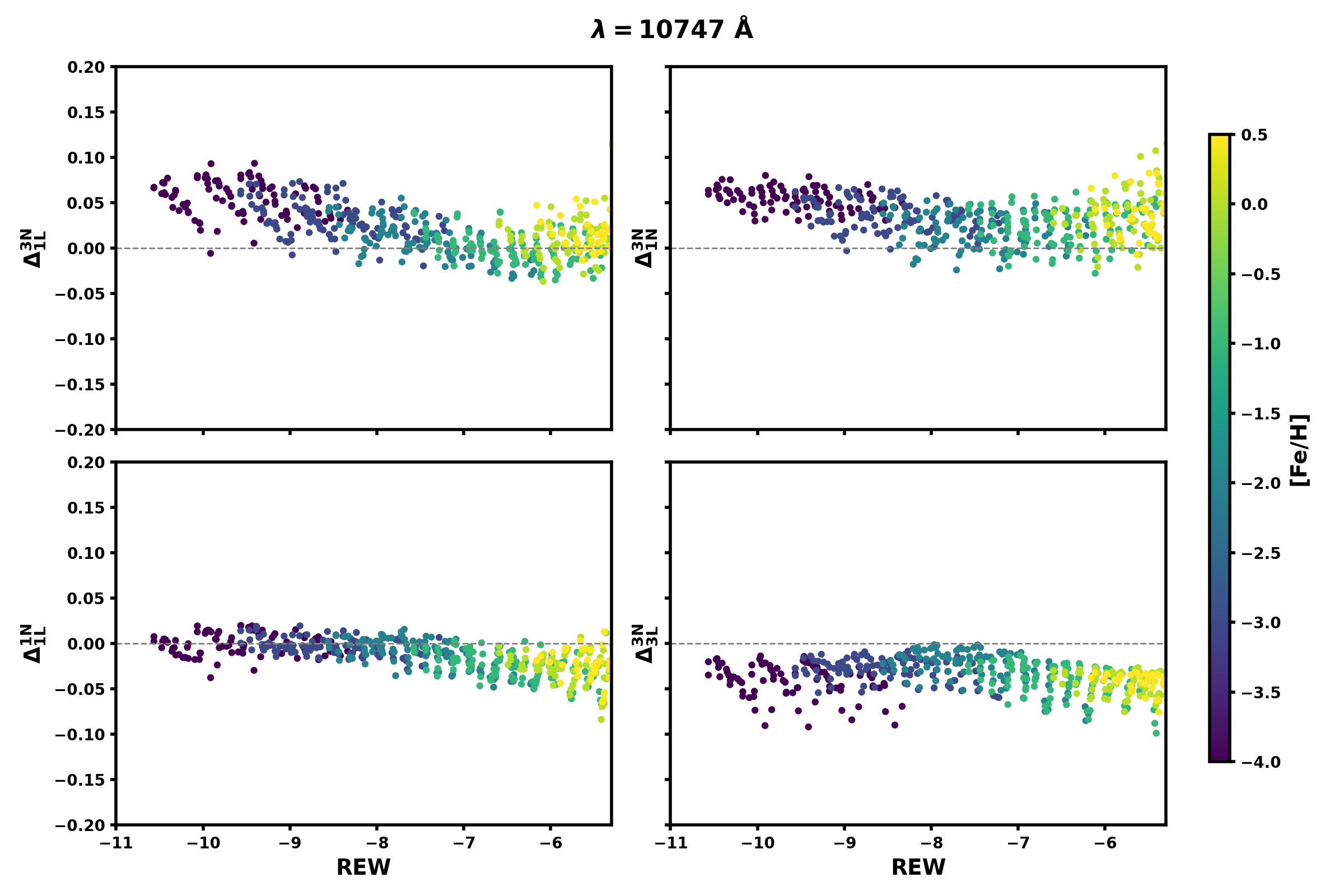}
    \caption{Abundance corrections for the \ion{Na}{I} line at $10\,747$\,\r{A} between different models: 3D~NLTE to 1D~LTE (top left), 3D~NLTE to 1D~NLTE (top right), 1D~NLTE to 1D~LTE (bottom left), and 3D~NLTE to 3D~LTE (bottom right), colour-coded as their [Fe/H]. The datapoints are computed at [Na/Fe]$=-0.5$ to $+0.5$ for the dwarfs, and $+1.0$ for the giants, in steps of 0.5\,dex and, for the 1D models, at a $v_\mathrm{mic}=1.0$\,km s$^{-1}$.} 
    \label{fig: ALLabcorr10747}
\end{figure*}
\begin{figure*}
    \centering
    \includegraphics[width=1.0\linewidth]{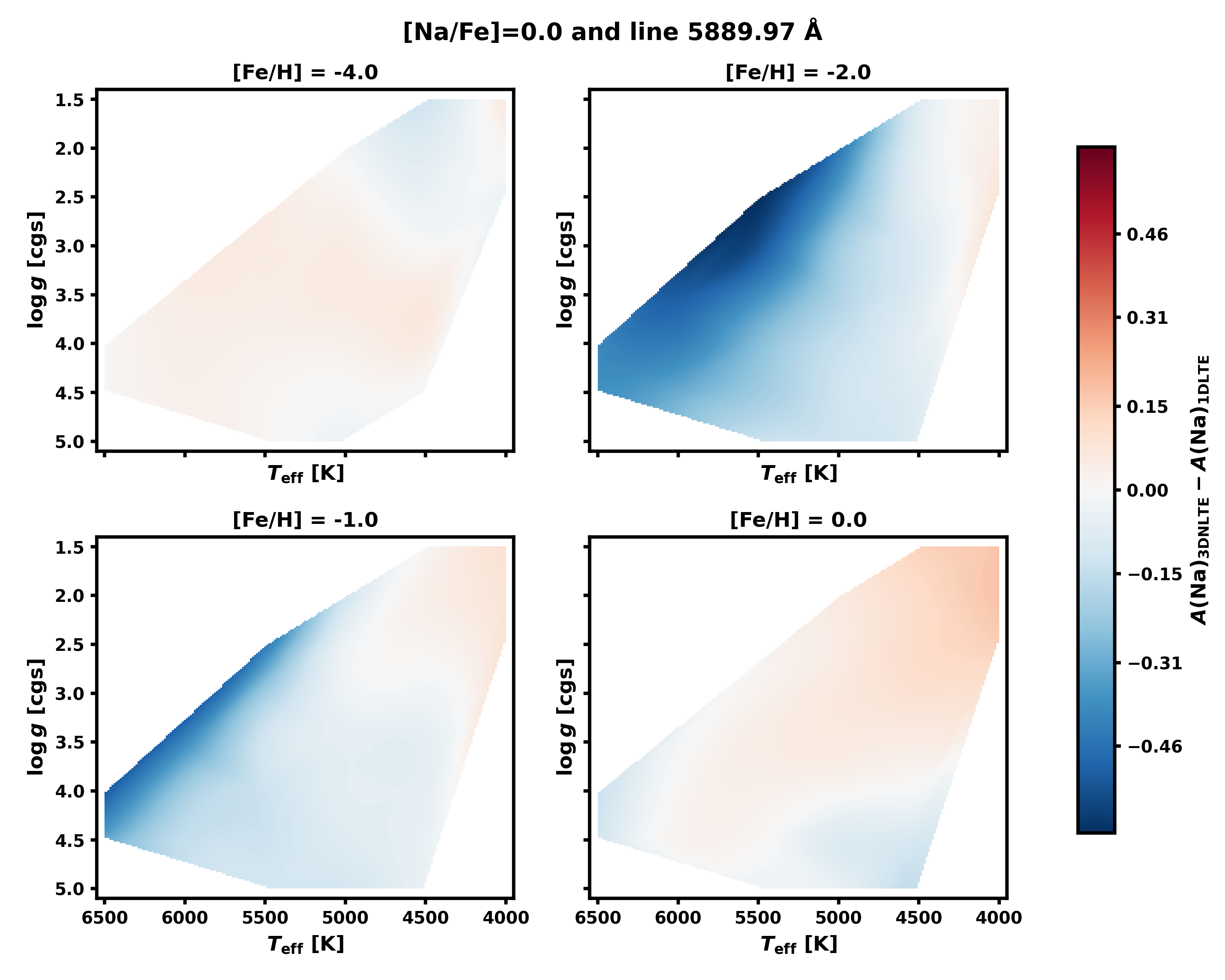}
    \caption{3D~NLTE - 1D~LTE abundance corrections ($\Delta^\mathrm{3N}_\mathrm{1L}$) for the 5889\,\r{A} line, shown at [Na/Fe]=0.0 and 1D models with $v_\mathrm{mic}=1.0$\,km\,s$^{-1}$. Each panel corresponds to a different metallicity, and specifically, from left to right, top to bottom: [Fe/H]$=-4.0, -2.0, -1.0, 0.0$.} 
    \label{fig: abcorr5889HR}
\end{figure*}

%
%





\end{document}